\documentclass[fleqn,usenatbib]{mnras}

\usepackage{newtxtext,newtxmath}

\usepackage[T1]{fontenc}
\usepackage{ae,aecompl}

\usepackage{graphicx}	
\usepackage{amsmath}	
\usepackage{diagbox}
\usepackage{colortbl}
\definecolor{gray}{rgb}{0.6,0.6,0.6}

\newcommand{\dd}{\mathrm d}
\renewcommand{\d}{\partial}
\renewcommand{\i}{\mathrm i}
\renewcommand{\(}{\left(}
\renewcommand{\)}{\right)}
\renewcommand{\v}[1]{\boldsymbol{#1}}

\usepackage{soul}	
\usepackage[dvipsnames]{xcolor}

\title[Cosmic ray production in superbubbles]{Cosmic ray production in superbubbles}

\author[T. Vieu et al.]{
T. Vieu,$^{1}$\thanks{E-mail: vieu@apc.in2p3.fr}
S. Gabici,$^{1}$
V. Tatischeff,$^{2}$
S. Ravikularaman$^{1}$
\\
$^{1}$Universit\'e de Paris, CNRS, Astroparticule et Cosmologie, F-75013 Paris, France\\
$^{2}$Universit\'e Paris-Saclay, CNRS/IN2P3, IJCLab, 91405 Orsay, France\\
}

\date{Accepted XXX. Received YYY; in original form ZZZ}

\pubyear{2019}

\begin{document}
\label{firstpage}
\pagerange{\pageref{firstpage}--\pageref{lastpage}}
\maketitle

\begin{abstract}
We compute the production of cosmic rays in the dynamical superbubble produced by a cluster of massive stars. Stellar winds, supernova remnants and turbulence are found to accelerate particles so efficiently that the nonlinear feedback of the particles must be taken into account in order to ensure that the energy balance is not violated. High energy particles do not scatter efficiently on the turbulence and escape quickly after each supernova explosion, which makes both their intensity inside the bubble and injection in the interstellar medium intermittent. On the other hand, the stochastic acceleration of low energy particles hardens the spectra at GeV energies. Because cosmic rays damp the turbulence cascade, this hardening is less pronounced when nonlinearities are taken into account. Nevertheless, spectra with hard components extending up to 1 to 10 GeV and normalised to an energy density of 1 to 100~eV~cm$^{-3}$ are found to be typical signatures of cosmic rays produced in superbubbles. Efficient shock reacceleration within compact clusters is further shown to produce hard, slightly concave spectra, while the presence of a magnetised shell is shown to enhance the confinement of cosmic rays in the bubble and therefore the collective plasma effects acting on them. We eventually estimate the overall contribution of superbubbles to the galactic cosmic ray content and show typical gamma-ray spectra expected from hadronic interactions in superbuble shells. In both cases, a qualitative agreement with observations is obtained.
\end{abstract}

\begin{keywords}
acceleration of particles -- shock waves -- cosmic rays
\end{keywords}



\section{Introduction}
Superbubbles (SBs) are expanding cavities carved inside dense molecular clouds by the feedback of clustered massive stars. It is thought that most massive stars are born in either compact or loose associations, in which they spend most of their lives until they explode as supernovae \citep{higdon2005}.
On the other hand, massive stars have long been thought to be the major sources of galactic cosmic rays (CR), in particular via diffusive shock acceleration (DSA) at wind termination shocks (WTS) and supernova remnants (SNR) \citep[e.g.][]{cesarsky1983,gabici2019}.

Inside SBs, star winds and supernovae combine to inject mechanical energy in the surrounding medium. It was early realised \citep[e.g.][]{montmerle1979} that this energy could be available for efficient particle acceleration, in particular due to collective plasma effects \citep{bykov1992b}. This scenario is an alternative to the standard paradigm of CR acceleration in isolated SNR. 
A number of phenomenological studies \citep [e.g.][]{higdon1998,parizot1999,higdon2006,tatischeff2018} support the SB origin of CRs. The composition of galactic CRS also points to an origin of these particles in SBs \citep{tatischeff2021}. Besides, particle reacceleration at multiple shocks and strong turbulence inside large superbubbles is expected to be more efficient than the acceleration around single supernovae, such that one can hope to push the maximum energy above the spectral ``knee'', as required to explain the transition between the galactic and extragalactic components of the CR spectrum measured in the vicinity of the Earth. Although this scenario for the origin of CRs has been explored over the years \citep[see the reviews by][and references therein]{parizot2004,bykov2014,lingenfelter2018,bykov2020}, a renewed interest in it rised after the recent detections in gamma-rays of star clusters and SBs, both in the Milky Way (MW) and in the Large Magellanic Cloud (LMC) \citep{aharonian2007,abramowski2012,ackermann2011,abramowski2015,katsuta2017,aharonian2019}. This confirms that stellar clusters and SBs are important CR sources. However, few attempts have been made to model them in a self-consistent way.

\citet{klepach2000} considered CR acceleration either at the collective termination shock produced by the star winds in a compact cluster, or around the multiple individual winds in the case of a loose cluster. They found that the maximum CR energy could be increased by up to two orders of magnitude in these systems compared to the acceleration at a single shock, and that the more efficient acceleration would lead to hard spectra. Efforts to model the acceleration of particles in interacting shocks, such as colliding winds or SNRs, have been made in a series of paper by Bykov and collaborators \citep{bykov2013,bykov2018}. It was shown that very hard spectra could be expected in the idealised case of stationary plane infinite shocks. An attempt to include geometrical effects has been performed in \citet{bykov2015}, and an insight in the time-dependent aspects was discussed in \citet{vieu2020}.
Models of CR acceleration at the collective wind of a compact cluster were developed recently by \citet{gupta2020} and \citet{morlino2021}. Both studies showed that the maximum energy of accelerated protons could exceed 1~PeV if the magnetic field close to the termination shock is efficiently amplified, which is required to better confine the particles around the shock. The stochastic acceleration in the turbulent medium was not included in these models, even though a substantial fraction of the stellar energetic input is believed to be stored in turbulent MHD waves \citep[e.g.][]{butt2008,gallegosgarcia2020}.

In the 90's, Bykov et al. used renormalisation methods \citep{bykov1993} to model the acceleration of particles inside young and turbulent SBs obtaining interesting results regarding the shape of the spectrum and break energies \citep{bykov1995,bykov2001}. This stochastic model includes the effect of the strong, possibly supersonic, turbulence, and also accounts for the particle feedback onto turbulence, making it the first - and up to now, to our knowledge, only - self-consistent modelling of CR acceleration in SB environments. A time-dependent model was later developed by \citet{ferrand2010}, who considered the reacceleration of the particles by successive SNR shocks sweeping the entire SB volume, as well as the modulation of the spectrum in weak turbulence. However, the model of \citet{ferrand2010} neglected the backreaction of the accelerated particles on the plasma and assumed that stellar winds and energy losses would provide negligible contributions to the nonthermal particle population. More recently, \citet{tolksdorf2019} solved the CR transport around SB cavities. They showed that interstellar particles could be efficiently reaccelerated in the turbulent interior. The primary acceleration at shock waves was however disregarded.

In the present work, we seek to develop a self-consistent model of CR acceleration in SBs, including all the relevant physical ingredients, namely stellar winds, SNRs, turbulence, energy losses, and also accounting for the backreaction of the particles on the shocks and the turbulence. Section~\ref{sec:2} discusses the SB environment. The acceleration mechanisms are modelled in Section~\ref{sec:3}. Section~\ref{sec:results} details the main results regarding the CR energetics, intermittency and spectra inside a typical SB. The model is then refined in Section~\ref{sec:5} by considering two different regions, which can either account for a compact cluster inside the SB or for the presence of a magnetised shell around the SB. We estimate the SB contribution to the galactic CRs in Section~\ref{sec:6} and compute typical gamma-ray signatures in Section~\ref{sec:7}. We eventually conclude in Section~\ref{sec:conclusions}.

In order to improve the clarity of the paper, a summary of the notations used throughout this work is provided in Table~\ref{tab:notations}.


\section{Superbubble properties}\label{sec:2}

\subsection{Bubble structure and stellar power}\label{sec:cluster}
\begin{figure}
   \centering
   \includegraphics[width=\linewidth]{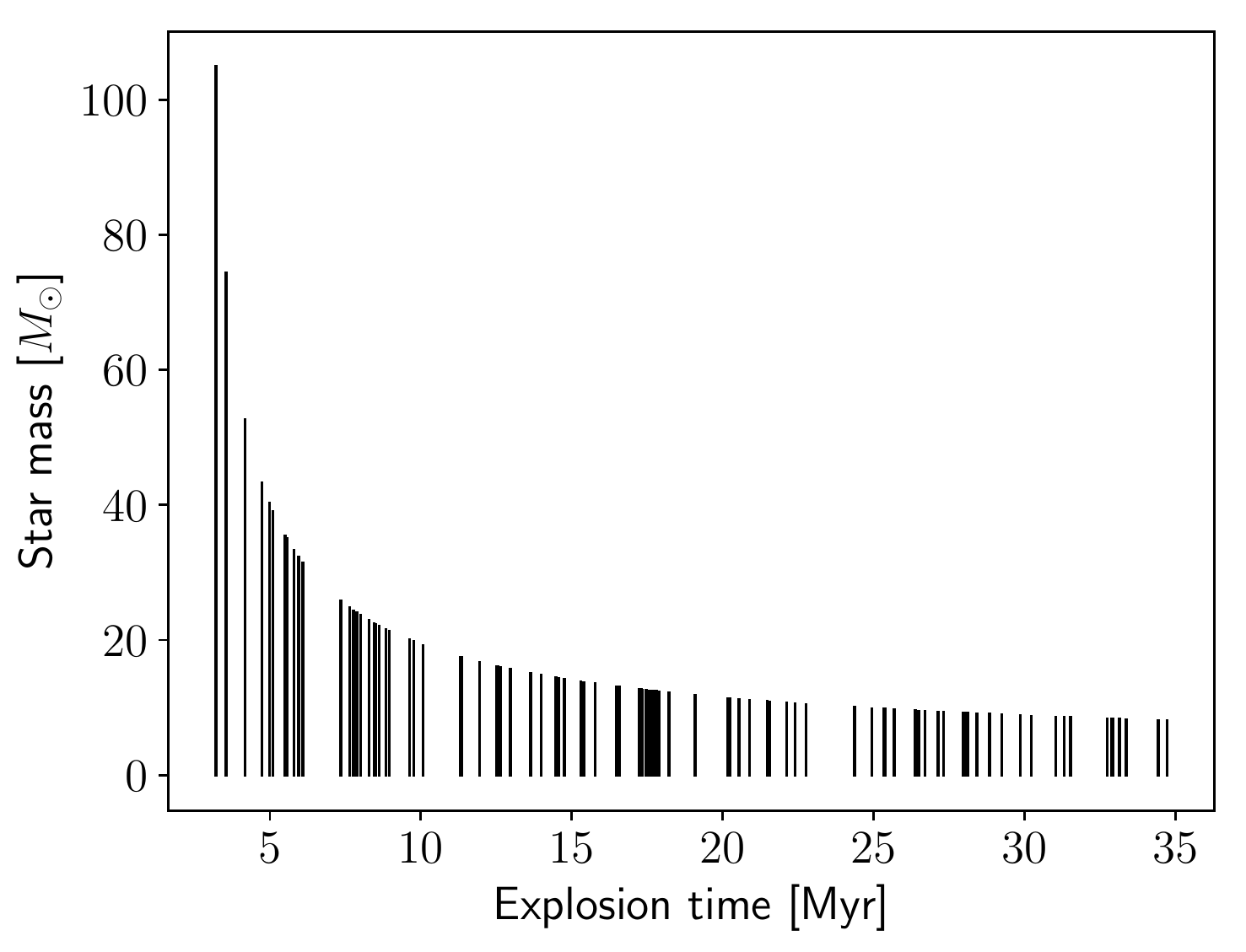} 
  \caption{Supernova explosion times as a function of stellar mass for a cluster of 100 stars in the range 8-150 $M_{\odot}$.}
  \label{fig:listcluster}
\end{figure}

We consider a cluster of hundreds of massive ($8-150~M_\odot$) stars distributed according to an initial mass function of index 2.3 \citep{salpeter1955}. The mass of a star can be connected to its lifetime by using the numerical fit from the stellar evolution simulations performed by \citet{limongi2006}: $\log_{10} (t ~\textrm{[yr]}) = 9.598-2.879\log_{10}M+0.6679(\log_{10}M)^2$, where $M$ is the initial mass of the star in solar masses. Assuming that all stars in a cluster are born at the same time $t = 0$, it is possible to implement a Monte Carlo sampling to simulate the times at which they explode into supernovae. One of such realisations for a cluster of 100 massive stars is shown in Figure~\ref{fig:listcluster}. Such a number of massive stars represents the typical content of superbubbles \citep{lingenfelter2018}. The most massive stellar clusters located near the galactic centre contain of the order of $10^3$ massive stars \citep{krumholz2019}.

It should be noted that supernova explosions are still poorly understood. In particular, simulations suggest that very massive stars (> 20-40~$M_\odot$) may only produce a weak, sometimes called \textit{failed} supernova \citet{fryer1999,fryer2003}, while stars with masses higher than 40~$M_\odot$ may not explode at all and directly collapse onto a black hole.
There is still no consensus nowadays about the fate of massive stars. 
It may well be the case that only a narrow mass interval allows for a powerful supernova explosion \citep{sukhbold2016,ebinger2020}.
Disregarding these debated issues, we assume in the following that all massive stars, i.e. with mass higher than 8~$M_\odot$, explode in a powerful supernova. In a typical cluster characterised by a Salpeter initial mass function, only 25\% of the stars have a mass higher than 20~$M_\odot$, so the controversial fate of these stars does not affect significantly the mechanisms of CR acceleration.

Stars inject kinetic energy in the surrounding medium not only when they explode as supernovae, but also during their entire lives, mainly due to the presence of stellar winds \citep{cesarsky1983,seo2018}. A cavity filled by hot and diluted gas, called SB, forms as the result of the combined effect of stellar winds and supernova explosions. A simple dimensional argument can be invoked to show that the radius of the bubble increases with time as \citep{castor1975,weaver1977,mccray1987}:
\begin{align}
\label{eq:weaver}
R_b(t)
&\approx 68~ \xi_b^{1/5} \left( \frac{L_*}{10^{38}~{\rm erg/s}}\right)^{1/5} \left( \frac{n_{\rm ISM}}{\rm cm^{-3}} \right)^{-1/5} \left( \frac{t}{\rm Myr} \right)^{3/5} \rm pc
\, ,
\end{align}
where $L_*$ is the total average mechanical power injected in the SB by stellar winds and supernova explosions, $n_{\rm ISM}$ is the hydrogen number density of the surrounding medium, and $\xi_b$ a numerical factor to be determined from more accurate studies \citep[e.g.][]{yadav2017}. In the expression above, physical quantities have been normalised to typical values, and the implicit assumptions have been made that $R_b$ is much larger than the radius of the star cluster and that SNRs become subsonic before reaching the edge of the SB.

To estimate the wind mechanical power, denoted $\mathcal{P}_w$ in the following, we account for both the main sequence (MS) and Wolf-Rayet (WR) wind phases, while we neglect the red supergiant phase which gives a negligible contribution \citep{seo2018}.
Stellar evolution models provide an estimate of the wind power in either phase as \citep{seo2018}:
\begin{align}\label{windpowr}
\begin{aligned}
    \log_{10} \( \frac{\mathcal{P}_{w,MS}(M)}{\mathrm{~erg~Myr}^{-1}}\) &\approx -3.4 \( \log_{10} M  \)^2 + 15 \log_{10} M +34 \, ,
\\
    \mathcal{P}_{w,WR}(M)&\approx
    6 \times 10^{48} M^{1.2} \mathrm{~erg~Myr}^{-1} \, .
\end{aligned}
\end{align}
While the duration of the MS is computed from the fits provided by \citet{limongi2006}, the duration of the WR phase is assumed to be 320~kyr, independently on the initial mass of the star provided it is higher than 20~$M_\odot$ \citep{seo2018}.
 
As an example, the lifetime of a star of initial mass $M\sim 20~M_\odot$ is around 20 Myr, its total MS wind energy is of the order 10$^{49}$~erg, and its total wind energy in the WR phase is around 10$^{50}$~erg. For the most massive O stars, $M\sim 100~M_\odot$, the total wind energy reaches  10$^{51}$~erg, which is comparable to the mechanical energy of a supernova explosion.

\begin{figure}
   \centering
  \includegraphics[width=\linewidth]{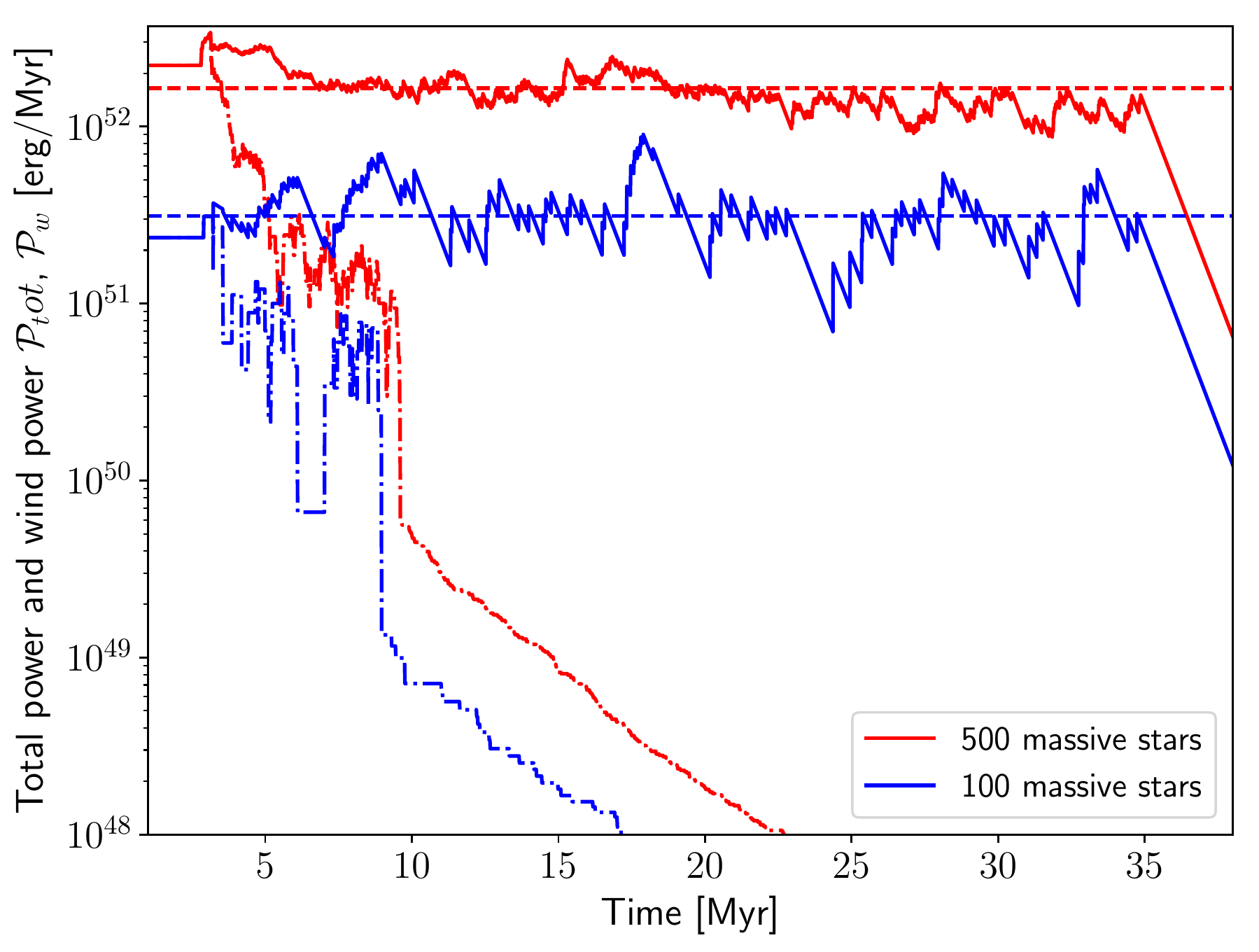}
  \caption{Evolution of the mechanical power of the cluster assuming a relaxation time $\tau_{T} = 1$~Myr for the SNR perturbations (see Equations~\ref{windpowr} and~\ref{CRinSB_Sourceturbulence}). ``Massive'' stars are stars with $M > 8 M_\odot$. The dash-dotted lines indicate the contribution of the winds, $\mathcal{P}_w$. The dashed lines indicate the average mechanical power.}
  \label{fig:evolpower}
\end{figure}

The mechanical power of two randomly generated clusters containing respectively 100 and 500 massive stars is plotted in Figure~\ref{fig:evolpower}. In the early stage of the cluster ($t \lesssim 5$~Myr), the contribution of the WR stars dominates, while in older SBs the supernovae are the main sources of mechanical power. Because the WR and SNR phases last for a short time, the injection of mechanical energy in the bubble is intermittent, although the mean value is roughly constant in time. The wind power quickly drops after 10~Myr, which is the time at which all stars of mass higher than $20~M_\odot$ have exploded. Then there only remain MS stars. Eventually, the average power of supernovae over the lifetime of the cluster is $9 \times 10^{35}$~erg/s/star while the average power of the winds of massive ($M > 8 M_\odot$) stars is $9 \times 10^{34}$~erg/s/star, for a total average power of $10^{36}$~erg/s/star. The winds are therefore subdominant, yet non negligible, in particular because the input from supernovae is strongly intermittent. The total production of energy by a cluster of 100 massive stars is of the order $10^{53}$~erg. Note that the average power injected into the SB, $\int_0^{\mathcal{T}_{\rm SB}} \dd t \mathcal{P}_{\rm tot}(t) / \mathcal{T}_{\rm SB}$, where $\mathcal{T}_{\rm SB}$ is the SB lifetime, is equal to the average luminosity (or mechanical power) of the cluster $L_*$ which drives the dynamics of the bubble.
Eq.~\ref{eq:weaver} can therefore be rewritten as:
\begin{equation}
    R_b(t) \approx R_0 N_*^{1/5} \left( \frac{t}{\rm Myr} \right)^{3/5}
    \, , 
\end{equation}
where $N_*$ is the number of massive ($M > 8 M_\odot$) stars in the cluster and $R_0 = 27~ \xi^{1/5} \left( n_{\rm ISM}/\rm cm^{-3} \right)^{-1/5} \rm pc$. Assigning a numerical value to $R_{0}$ is sufficient to determine the entire dynamical evolution of the SB.

\begin{figure}
\centering
  \includegraphics[width=1.\linewidth]{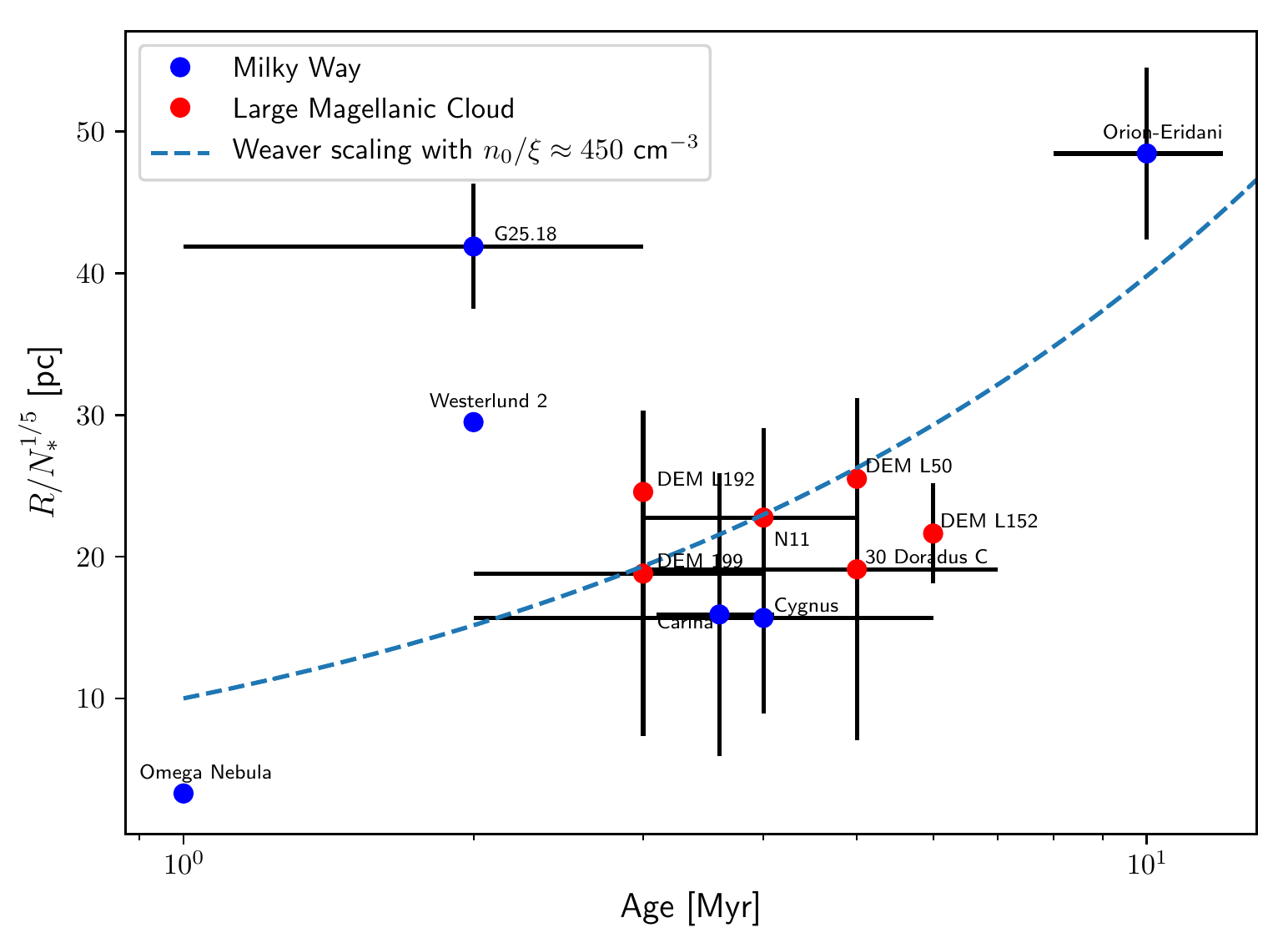}
\caption{Sizes and ages of several observed clusters. The error bars reflect the variations found in the literature rather than the physical uncertainties. In particular, the number of massive stars in the G25 region is only roughly estimated, and Westerlund 2 is poorly constrained.
Data are found in \citet[and references therein]{ackermann2011} (Cygnus); \citet[and references therein]{joubaud2019},\citet{voss2010} (Orion-Eridani); \citet[and references therein]{lopez2020} (30 Doradus C); \citet[and references therein]{katsuta2017} (G25); \citet{maddox2009} (N11); \citet{dunne2003} (Omega nebula); \citet{smith2006,smith2007} ($\eta$-Carina); \citet{rauw2007} (Westerlund 2); \citet{cooper2004} (DEM L192); \citet[and references therein]{jaskot2011} (DEM L152 and DEM L50); \citet{dunne2001,massey1995} (DEM 199). See also the compilation in \citet[and references therein]{ferrand2010}.}
\label{fig:plotweaver}
\end{figure}

Figure~\ref{fig:plotweaver} displays the radius of several known SBs as function of their age. These observations suggest the scaling $n_{\rm ISM}/\xi_b \approx 450$~cm$^{-3}$.  In the following we will assume $n_{\rm ISM} = 100$~cm$^{-3}$, which is a typical density of giant molecular clouds where young clusters are expected to be born and expand \citep{parizot2004}. This implies $\xi_b \approx 22$\% and $R_0 \approx 10$~pc.

As the SB expands, the medium is diluted and cooled. The seminal SB model of \citet{weaver1977} provides the following self-similar scalings for the average density and temperature inside the bubble \citep[][]{maclow1988}:
\begin{align}
    n(t) &= 0.34  \( \frac{\xi_b N_*}{100} \)^{\frac{6}{35}} \( \frac{n_{\rm ISM}}{100~ {\rm cm}^{-3}} \)^{\frac{19}{35}} \( \frac{t}{{\rm Myr}}\)^{-\frac{22}{35}} ~{\rm cm}^{-3} \, ,
    \\
    T(t) &= 4.8 \times 10^6  \( \frac{\xi_b N_*}{100} \)^{\frac{8}{35}} \( \frac{n_{\rm ISM}}{100~ {\rm cm}^{-3}} \)^{\frac{2}{35}} \( \frac{t}{{\rm Myr}}\)^{-\frac{6}{35}} ~{\rm K}
    \, .
\end{align}
Together with the expansion of the outer shock, the evolution of the density and the temperature describes the dynamical SB environment.

\subsection{Turbulence generation}
Inside the SB, the stellar mechanical energy is converted into turbulence which cascades from the largest scale to the dissipation scale. We assume equipartition of the kinetic and magnetic energies in the waves and denote $W$ the spectral energy density such that $\int \dd k W(k) = \delta B^2/4 \upi$. The dynamics of the turbulence can be described by the following nonlinear local energy transfer equation \citep{zhou1990,norman1996,miller1995,ptuskin2005}:
\begin{equation}\label{turbulence_cascade}
    \d_t W + \d_k \( \frac{a}{\sqrt{\rho}} k^{5/2} W^{3/2} \) = - \Gamma(k) W + S \delta(k-k_0) \, ,
\end{equation}
where $a \approx 0.8$ is a numerical constant determined from experiments or simulations \citep[e.g.][]{verma1996}, $\rho$ is the gas density, $2\upi/k_0$ is the largest turbulent scale, $S$ is the injection of energy at the largest scale and $\Gamma$ is the damping rate.

We assume that both winds and supernovae convert a fraction $\eta_T$ of their mechanical energy into magnetised waves, and that the conversion is exponentially suppressed for SNRs, such that the source term of the turbulence cascade is phenomenologically written as:
\begin{align}\label{CRinSB_Sourceturbulence}
\begin{aligned}
    S &= \frac{\eta_T}{V_{\rm SB}(t)} \mathcal{P}_{\rm tot} \, ,
    \\
    \mathcal{P}_{\rm tot} &= \sum_{i=1}^{N(t)} \mathcal{P}_w(M_i) + \sum_{N(t)+1}^{N_*} \frac{E_{\rm SN}}{\tau_T} e^{-(t-t_i)/\tau_T}
    \, ,
\end{aligned}
\end{align}
where $N(t)$ is the number of stars remaining in the cluster at time $t$, $t_i$ is the time at which the star number $i$ is expected to explode, $E_{\rm SN} = 10^{51}$~erg is the energy released per supernova and $\tau_T$ is the relaxation time of the turbulence after a supernova explosion.
This recipe is similar to that adopted in \citet{fang2019} for the conversion of the wind power. The exponential decay of the turbulence excited by SNRs is motivated by numerical simulations \citep[e.g.][]{krause2013}, which suggest a relaxation time $\tau_T \approx 1$~Myr.

In the stationary regime, the general solution of Eq.~\ref{turbulence_cascade} reads:
\begin{align}\label{solution_turbulence}
    W(k) &= k^{-5/3} \( \frac{\sqrt{\rho} S}{a} \)^{2/3} \( 1 - \frac{1}{3} \( \frac{\rho}{S a^{2}} \)^{1/3} \int_{k_0}^k \dd k' \frac{\Gamma(k')}{k'^{5/3}} \)^2 \, .
\end{align}
In the absence of damping, one retrieves the Kolmogorov scaling $W \propto k^{-5/3}$. The generalisation to another turbulence regime (e.g. Kraichnan) is straightforward.
The test-particle solution provides an estimate of the energy density contained in the turbulence:
\begin{equation}\label{CR_in_SBturbulenceenergy}
    \frac{\delta B^2}{4 \upi} \approx \rho \delta u^2 \approx \frac{3}{2} \( \frac{\sqrt{\rho} S}{a k_0} \)^{2/3} \, .
\end{equation}
Considering turbulence strengths above this limit would violate energy conservation in the cluster. Eq.~\ref{CR_in_SBturbulenceenergy} allows to estimate the turbulent magnetic field $\delta B$ and velocity $\delta u$:
\begin{align}\label{estimate_dB}
    \delta B &\lesssim 5 \( \frac{n}{0.01 ~{\rm cm}^{-3}} \)^{\frac{1}{6}} \(
     \frac{\eta_T \mathcal{P}_{\rm tot}}{10^{51} ~{\rm erg/Myr}} \frac{\lambda}{10~{\rm pc}} \)^{\frac{1}{3}} \( \frac{R_b}{50~{\rm pc}} \)^{-1} ~\text{\textmu G}
     \, ,
     \\
     \delta u &\lesssim 110 \( \frac{n}{0.01~{\rm cm}^{-3}} \)^{-\frac{1}{3}} \(
     \frac{\eta_T \mathcal{P}_{\rm tot}}{10^{51} ~{\rm erg/Myr}} \frac{\lambda}{10~{\rm pc}} \)^{\frac{1}{3}} \( \frac{R_b}{50~{\rm pc}} \)^{-1} ~\text{km/s}
     \, ,
\end{align}
where $\lambda = 2 \upi/k_0$ is the largest turbulent scale.

The density, power and radius of the SB depend on the total number of stars $N_*$ as well as the time $t$. From the scalings derived above ($n \propto N_*^{6/35} t^{-22/35}$; $\mathcal{P}_{\rm tot} \propto N_*$; $R_b \propto N_*^{1/5} t^{3/5}$), we get to conclude that the random component of the velocity $\delta u$ is weakly dependent on the total number of stars: $\delta u \propto N_*^{8/105}$, while it decreases with time as $\delta u \propto t^{-41/105}$. The sound speed being of the order of 100 km/s inside the superbubble, the turbulence is expected to consist in an ensemble of weak secondary shock waves in the early times of the cluster history, namely the first few million years.
The transport of charged particles in such strong supersonic turbulence has been investigated by \citet{bykov1990,bykov1993}, who later showed that the acceleration of the particles could be very efficient, with up to 30 \% of the energy of the waves converted into CRs and such that hard CR spectra should be expected up to TeV energies \citep{bykov1992a,bykov1992b,bykov2001}, while a steep component could remain up to energies as high as 100 PeV \citep{bykov1995}. In the following we shall rather analyse the acceleration of particles in more evolved SBs, in which the turbulence is expected to be subsonic. One should nevertheless keep in mind that the early phase of efficient particle acceleration in supersonic turbulence could last several million years and be an efficient CR accelerator \citep[e.g.][]{bykov2001}.

\begin{figure}
   \centering
   \includegraphics[width=\linewidth]{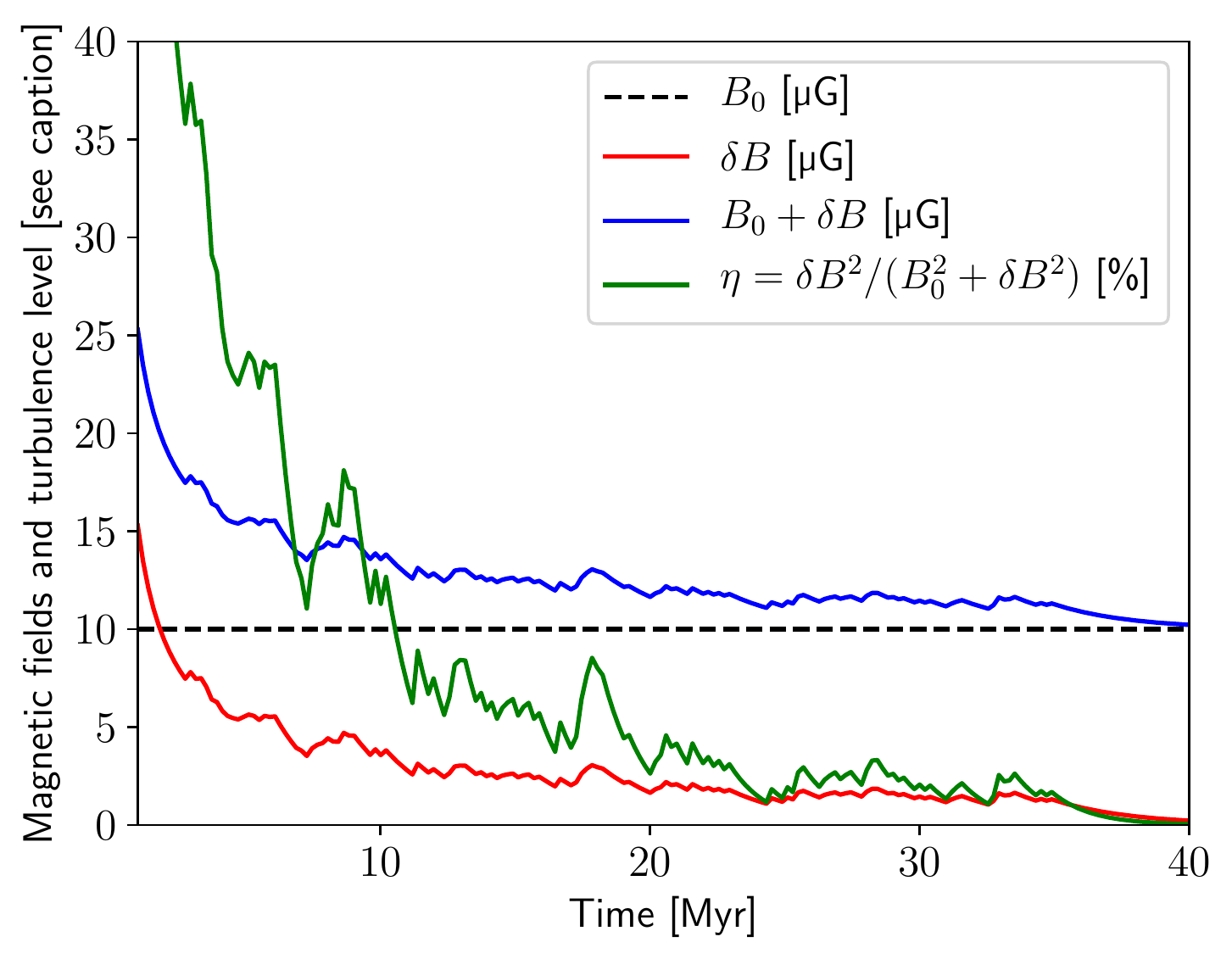} 
  \caption{Evolution of the turbulent magnetic field and turbulence level in a typical SB with $N_*=100$, $B_0 = 10$~\textmu G, $\eta_T = 30\%$, as estimated from Eq.~\ref{CR_in_SBturbulenceenergy}.}
  \label{fig:evolution_dB}
\end{figure}

The evolution of the random component of the magnetic field is plotted in Figure~\ref{fig:evolution_dB} for a cluster containing 100 massive stars, where we assumed a background field $B_0 = 10$~\textmu G and an efficiency of turbulence generation $\eta_T = 30 \%$. The quasi-linear theory is, strictly speaking, only valid when $\delta B < B_0$, or $\eta < 50 \%$, where $\eta = \delta B^2/(B_0^2 + \delta B^2) $ is the turbulence level. This is verified after a few million years for a background field $B_0 = 10$~\textmu G. On the other hand, the turbulence may remain strong until the end  of the cluster life if e.g. $B_0 = 1$~\textmu G (still assuming $\eta_T = 30 \%$).

The turbulence level decreases rapidly in the first million years as WR stars explode and the wind power diminishes. Then it stabilises after about 10~Myr around a few percent (for $B_0 = 10$~\textmu G), when only MS stars remain and the main sources of turbulence are the SN explosions.

\section{Particle acceleration and transport in superbubbles}\label{sec:3}
\subsection{Particle acceleration at stellar wind termination shocks}
The stellar outflows may form a collective wind termination shock (WTS) around a star cluster. The position of the WTS $R_s$ is determined by the equilibrium between the ram pressure of the collective wind and the pressure in the interior of the SB, which provides \citep[e.g.][]{gupta2020,morlino2021}:
\begin{multline}\label{CR_in_SB:radiusTS}
R_s \approx \frac{(\xi_b L_*)^{-1/5} \mathcal{P}_w^{1/2} }{\(10^{36}~{\rm erg/s} \)^{3/10} } \( \frac{n_{\rm ISM}}{100~{\rm cm}^{-3}} \)^{-3/10} \(\frac{V_w}{2000~{\rm km/s}} \)^{-1/2} \\ \times \(\frac{t}{1~{\rm Myr}} \)^{2/5} ~{\rm pc}
\, ,
\end{multline}
where $L_* \approx 10^{36} N_*$~erg/s is the average mechanical power imparted by the winds and the supernovae (see the dashed lines in Fig.~\ref{fig:evolpower}), which determines the pressure in the SB interior, while $\mathcal{P}_w$ is the mechanical power of the winds, which determines the ram pressure of the WTS. Eq.~\ref{CR_in_SB:radiusTS} is only valid when the wind power is independent of time. Yet it provides a reliable estimate of the radius of the termination shock in our setup, where the wind power is constant most of the time and varies only significantly during the brief WR phase of the stars just before the SN explosions. In the early phase of the cluster life, the WTS radius may reach about 10~pc. On the other hand, after 10~Myr only MS stars remain and the wind power is subdominant, such that the collective WTS can barely be sustained and shrinks to sub-parsec sizes. For simplicity, we will therefore consider \textit{loose clusters} in the following, assuming that the extent of the cluster (or that of the region of energy deposition) is of the order of 10~pc. In this case, a collective WTS cannot form and the winds produce isolated strong shocks around each stars, with radii determined by Eq.~\ref{CR_in_SB:radiusTS} provided the total wind power $\mathcal{P}_w$ is replaced by the power of the individual stars.

A fraction of the mechanical energy of the winds can be converted into CRs via DSA operating at the individual WTS \citep{cesarsky1983,seo2018}. Because winds do not produce a high energy density of particles, we assume that this acceleration process takes place in the test-particle regime. We can therefore describe the acceleration of particles at the WTS as a continuous and time dependent injection of CRs characterised by an injection rate of particles of momentum $p$ equal to:
\begin{equation}\label{qwindinjection}
    q_w(p) = \frac{\eta_w p^{-4} e^{-p/p_w} }{\mathcal{N}_w} \sum_{i=1}^{N(t)} \mathcal{P}_w(M_i) \, ,
\end{equation}
with $\eta_w \approx 10 \%$ the fraction of the wind mechanical energy that goes into CRs, $\mathcal{P}_w(M_i)$ the mechanical power of the wind ejected by the $i$th star in the cluster, $p_w$ the maximum momentum of the particles, and $N(t)$ the number of stars remaining in the cluster (i.e. not yet exploded) at time $t$. The maximum momentum achievable in a nearly stationary WTS is expected to be limited by the size of the shock rather than its lifetime: $p_w \sim Z e B V_w R_s \sim 10^5$~GeV \citep[e.g.][]{voelk1988,morlino2021}.
Finally, $\mathcal{N}_w$ is a normalisation constant given by the following integral:
\begin{align}
    \mathcal{N}_w = 4 \upi \int_{p_0}^{p_w} \dd p \, \epsilon(p)  p^{-2} e^{-p/p_w} 
    ,
\end{align}
where $p_0$ is the injection momentum and $\epsilon(p)$ the kinetic energy of protons of momentum $p$.

\subsection{Particle reacceleration at supernova remnant shocks}\label{sec:CR_in_SB_reaccbySNR}
When a star explodes into a supernova, a SNR is formed. The SNR not only injects fresh CRs in the stellar cluster, but it also reaccelerates a fraction of the already existing particle distribution. The lifetime of a SNR in the medium of density $0.01 - 0.1$~cm$^{-3}$ is of the order $10 - 100$~kyr, which is short compared to the CR diffusion time away from the cluster (Myr), hence we assume that the SNR are instantaneous and compute CR propagation only between them \citep[as in][]{ferrand2010}. The CR spectrum after the passage of the SNR, $n_f(p)$, is written as:
\begin{equation}\label{reaccelerationSN}
    n_f = \mathcal{V}_{\rm SNR} \, \mathcal{R} \left[ \frac{n_i}{V_{\rm SB}} \right] + \(1- \frac{\mathcal{V}_{\rm SNR}}{V_{\rm SB}} \) n_i
    \, ,
\end{equation}
where $n_i(p)$ is the spectrum of CRs preexisting before the supernova explosion, $\mathcal{V}_{\rm SNR}$ is the volume spanned by the SNR, $V_{\rm SB} = 4 \upi /3 R_b^3 $ is the SB volume and $\mathcal{R}$ the ``reacceleration operator''. In the test-particle limit, this operator simply reads \citep[e.g.][]{melrose1993,ferrand2010}:
\begin{equation}
\mathcal{R} [f] = 4 \int_{p_0}^{r^{\frac{1}{3}} p} \frac{\dd p'}{p'} \( \frac{r^{\frac{1}{3}} p}{p'} \)^{-4} \(f(p') + \eta_{\rm SNR} \delta\(p'-p_0\)\) \, ,
\end{equation}
which describes both the injection of particles from the thermal bath with an efficiency $\eta_{\rm SNR}$ at the injection momentum $p_0$ and the reacceleration of the preexisting distribution of particles. However, the test-particle reacceleration operator cannot be applied in environments where multiple shocks are expected over small timescales, because this would systematically lead to the violation of the energy conservation as most SNR would transfer more than $10^{51}$~erg into non-thermal particles \citep[see][for a detailed discussion of this issue]{paper2}. The linear reacceleration operator must therefore be substituted by a nonlinear, self-consistent, computation.
In the following we use the procedure described in \citet{paper2}. In order to take into account the feedback of the CR pressure on the hydrodynamic profile and small-scale turbulence, two differential equations must be solved together with suitable boundary conditions. On one hand, the shock structure is modified by the non-thermal particles according to the conservation of momentum, while on the other hand the acceleration of the particles is driven by the transport equation integrated around the shock. The injection efficiency $\eta$ can be related to the injection momentum by imposing the continuity between the thermal and non thermal distributions of particles at the injection momentum \citep[][]{blasi2005}. Finally, adiabatic decompression is computed after the explosion of each supernova. As shown in \citet{paper2}, the spectrum of reaccelerated particles tends towards a universal shape after a few successive reaccelerations. 
Thus, this spectrum is expected to be largely independent of the injection parameter $\xi$, which is defined such that the injection momentum equals $\xi p_{\rm th}$, where $p_{\rm th}$ is the typical momentum of the thermal particles \citep[see][]{paper2}.
In the following, we will take $\xi=3.5$, which is chosen such that, in the absence of preexisting particles ahead of the shock, the remnant transfers about $10^{50}$~erg into non-thermal CRs. Assuming that SNR are strong shocks, the only remaining parameter is the maximum momentum. We constrain the latter in a simple and conservative approach, without assuming field amplification far upstream of the SNR.
The most stringent limitation on the maximum energy is therefore the acceleration time. The latter depends on the dynamics of the SNR, which is not straightforward to compute when the shock expands in an inhomogeneous medium.
Disregarding these complications, we assume that the velocity is constant in the free expansion phase, which is a good approximation especially when the remnant expands in a wind density profile \citep[e.g.][]{finke2012,gaggero2018}. SNRs expanding in a SB spend a large fraction of their life into the free expansion phase, at the end of which the mechanical power of the shock is maximal. The maximum momentum is also expected to be achieved at the end of the free expansion phase and can be estimated as:
\begin{align}
	p_{\rm max} = 1.4 \frac{Z B}{10~\text{\textmu G}} \(\frac{V_{\rm SN}}{3000 ~\text{km/s}}\) \( \frac{M_e}{10~M_\odot} \)^{\frac{1}{3}} \( \frac{n}{0.01 {{\rm cm}^{-3}}} \)^{-\frac{1}{3}} \text{~PeV}
	\, ,
\end{align}
where $n$ and $B$ are respectively the density and the total magnetic field in the SB interior and $Z$ is the atomic number of the particle.
Although this is admittedly a rough estimate, it allows to get rid of an otherwise free parameter without solving the time-dependent nonlinear problem of shock reacceleration accounting for the multiple environments the remnant experiences before entering the adiabatic phase. Besides, it contains all the relevant physical ingredients. The maximum momentum is indeed expected to increase with enhanced magnetic field and in low density media.

As shown in \citet{paper2}, it is expected that a substantial fraction of the non-thermal energy leaks upstream of the shock, carried by high-energy particles beyond the maximum momentum. Because these particles will quickly escape from the SB (within typically 10~kyr), they are not expected to affect the overall CR content and we disregard them in the present analysis. However, they should interestingly provide an intermittent input of very high energy particles in the Galaxy, leaving a signature in the region of the knee which is observed in the galactic cosmic ray spectrum. This additional component can however only be treated accurately using a time-dependent nonlinear model of CR reacceleration at SNR shocks, which is beyond the framework established in \citet{paper2} and will be considered in a future work.


The only remaining quantity to constrain is the volume spanned by the SNR, $\mathcal{V}_{\rm SNR}$, which determines the fraction of CRs reaccelerated by the expanding shock (Eq.~\ref{reaccelerationSN}). In the hot rarefied medium characterising the SB interior, a SNR becomes subsonic before becoming radiative \citep{maclow1988}. In principle, $\mathcal{V}_{\rm SNR}$ should therefore correspond to the volume spanned by the SNR before the forward shock weakens. However, because DSA is solved in the stationary regime the Mach number at which the shock ``weakens'' is not well defined. Instead of keeping it as a free parameter, we rather impose that the kinetic energy of the stationary strong shock multiplied by the volume $\mathcal{V}_{\rm SNR}$ equals the energy of the supernova explosion:
\begin{equation}\label{SNR_radius}
    \rho \mathcal{V}_{\rm SNR} u_0^2/2 = 10^{51}~{\rm erg}
    \, ,
\end{equation}
which ensures that the energy of the stationary shock will never exceed the energy budget of the time-dependent SNR.
The fraction of reaccelerated CRs thus reads:
\begin{equation}\label{fraction_reacc_CR}
    \frac{\mathcal{V}_{\rm SNR}}{V_{\rm SB}} \approx \min \( 0.59 \( \frac{R_b(t)}{20~\mathrm{pc}} \)^{-3} \( \frac{n(t)}{0.1 ~\mathrm{cm}^{-3}} \)^{-1} \( \frac{u_0}{1500 ~\mathrm{km/s}} \)^{-2}
    , 1 \)
    \, ,
\end{equation}
where $t$ is the age of the SB in Myr and $u_0 \approx 1500$~km/s is the velocity of the SNR shock at the beginning of the Sedov-Taylor phase, which is independent of the density and thus stays nearly constant throughout the SB evolution \citep{parizot2004}. Eq.~\ref{SNR_radius} could have been equivalently obtained by imposing that the acceleration process stops when the Mach number of the SNR shock falls below $\mathcal{M}_{\rm min} \approx 2.5 \left( \frac{u_0}{1500 ~\mathrm{km/s}} \right)^{7/2}  \left( \frac{T}{10^6~\rm K} \right)^{-1/2}$, which is consistent with the intuitive definition of a ``weak'' shock.
Assuming $N_*=100$ and $n_{\rm ISM} = 100$~cm$^{-3}$, and using the scaling described above, Eq.~\ref{fraction_reacc_CR} becomes: $\mathcal{V}_{\rm SNR}/V_{\rm SB} \approx 0.16 (t/{\rm Myr})^{-1.17}$, which means that only a few percent of the CR content will be reaccelerated by each SNR.

To summarise, in order to reduce an expanding supernova to a stationary shock, we assume that {\it (i)} the stationary shock velocity is equal to the velocity of the supernova at the beginning of the Sedov-Taylor phase ($u_0 = 1500$~km/s), {\it (ii)} the stationary shock disappears when its integrated kinetic energy $\rho \mathcal{V}_{\rm SNR} u_0^2/2$ is equal to the energy of the supernova explosion ($10^{51}$~erg), {\it (iii)} particles are injected from the thermal bath with a stationary efficiency such that in the absence of preexisting particles $10^{50}$~erg are transferred into CRs ($\xi = 3.5$), {\it (iv)} the time-independent maximum energy achieved by the freshly injected particles as well as that beyond which reaccelerated particles escape upstream is about that reached at the end of the free expansion phase\footnote{In principle the energy of the reaccelerated particles is not limited by the acceleration time but by the finite size of the shock. However both criteria provide very similar estimates at the beginning of the Sedov-Taylor phase.}, {\it (v)}~we disregard the upstream flux of very high energy particles assuming that they quickly leave the SB.

\subsection{Stochastic acceleration}
When they propagate inside the SB, particles experience resonant scattering against the magnetised waves. Each scattering deflects the particles in such a way that on average, an effective diffusion takes place both in space and momentum. The momentum diffusion operator reads:
\begin{equation}\label{diffusion in momentum}
    \d_t f = \frac{1}{p^2} \d_p \( p^2 D_p \d_p f(p) \) \, ,
\end{equation}
where $f(p)$ is the distribution function averaged over the spherical coordinates $\mu$ and $\phi$:
\begin{equation}
f(p) = \frac{1}{4 \upi} \int_0^{2 \upi} \dd \phi \int_{-1}^1 \dd \mu \, f(p,\mu,\phi)
\, ,
\end{equation}
with $\mu$ the pitch-angle cosine and $\phi$ the gyrophase, and $D_p$ is the pitch-angle averaged diffusion coefficient: $D_p(p) = 1/2 \int_{-1}^1 \dd \mu D_p(p,\mu)$. These normalisations ensure that the number density of the particles reads $n = 4 \upi \int \dd p p^2 f(p)$.

In the high energy regime $c \gg v_A$, with $v_A$ the Alfven speed, the quasi-linear theory of particle diffusion in turbulence provides the following expression for the pitch-angle averaged diffusion coefficient $D_p$ \citep{berezinskii1990}:
\begin{align}\label{diffusioncoeffQLT}
    D_p &= \frac{\upi^2 Z^2 e^2 v_A^2}{v} \int_0^1 \frac{\dd \mu}{\mu}\( 1- \mu^2 \) W\( \frac{Z e B}{\mu p} \) \, ,
    \\
    &= \frac{\upi^2 Z^2 e^2 v_A^2}{v} \int_{Z e B/p}^{\infty} \frac{\dd k}{k} \( 1- \( \frac{Z e B}{k p} \)^2 \) W\(k\) \, ,
\end{align}
where $W$ describes both co- and counter-propagating waves of equal energy, $Ze$ is the charge of the particle and $v$ its velocity. Here and in the following the background magnetic field $B_0$ is systematically supplemented by the perturbation $\delta B$ computed according to Eq.~\ref{estimate_dB} and we write the total field $B = B_0 + \delta B$. This allows to consistently account for the case of strong turbulence, $\delta B \gtrsim B_0$, where the quasi-linear theory only makes sense locally, for particles follow the mean resulting field. An accurate description of the strong turbulence regime is beyond the scope of this work. We recall that for the parameters considered in this analysis, the turbulence is strong only in the early phase of the SB evolution.

Once again, in the case where the energy of the diffusing particles is similar to that of the turbulence, one must account for the backreaction of the particles on the turbulence, which is expected to damp the waves. To include this nonlinear effect, we follow \citet{ptuskin2003,ptuskin2005,ptuskin2006,ptuskin2017}. Assuming an isotropic distribution of particles, the damping rate can be estimated through the following flux conservation \citep{eilek1979,thornbury2014}:
\begin{align}
    \int \dd k \Gamma(k) W(k)
    &= 4 \upi \int \dd p \, p^2 \frac{\epsilon(p)}{p^2} \d_p (p^2 D_p \d_p f) \, ,
    \\
    &=  4 \upi \int \dd p \, \d_p (v p^2 D_p) f(p) \, .
\end{align}
The quantity $\d_p (v p^2 D_p)$ is the mean energy transferred per unit time from the waves to the particles. Using Eq.~\ref{diffusioncoeffQLT}, we work out the right-hand side to identify:
\begin{equation}
\Gamma(k) = \frac{8 \upi^3 Z^2 e^2 v_A^2}{k} \int_{Z e B/k} \dd p \, p f(p) \, .
\end{equation}
This expression is identical to that written in e.g. \citet{eilek1979,miller1995,brunetti2004}. It differs from the expression appearing in the works of \citet[e.g.][]{ptuskin2006} by a factor $2 \upi$ due to the choice of normalisation of Eq.~\ref{turbulence_cascade}.

We assume that the turbulence cascade is faster than the other physical processes occurring in the SB such that we can neglect the time-dependency of the turbulence spectrum. We will comment a posteriori on the validity of this hypothesis.
From the stationary solution given by Eq.~\ref{solution_turbulence}, and assuming that the turbulence spectrum is steep enough for Eq.~\ref{diffusioncoeffQLT} to be approximated within a factor of order unity as
$D_p \approx 3 \upi^2 v_A^2 Z^2 e^2 W(Z e B/p)/5v$, we obtain:
\begin{multline}\label{diffusioncoeffwithfeedback}
    D_p 
    \approx \frac{3 \upi k_0^{1/3} p^{5/3} }{20 v} \(\frac{Z e B }{k_0}\)^{1/3} \( \frac{S}{a \rho} \)^{2/3} 
    \\
    \( 1 - \frac{2 \upi^2}{5} \( \frac{Z e B}{ \rho^2 a^2 S } \)^{1/3} \int_p \dd p' f(p') p'^{8/3} \)^2
    \, .
\end{multline}
%

The feedback of the particles on the turbulence is non trivial. Not only it reduces the density of the magnetic waves, but it can also terminate the turbulence cascade at a finite scale.
Indeed, it may happen that for high enough CR energy densities the quantity in the parenthesis in Eq.~\ref{diffusioncoeffwithfeedback} vanishes at small momenta. Low energy particles will not be able to resonate anymore with the magnetised waves, and their momentum diffusion coefficient should therefore be set to zero. This was already pointed out by \citet{ptuskin2006} in the context of CR diffusion in the interstellar medium.

Noticing that at $p=p_0$ the integral in the RHS of Eq.~\ref{diffusioncoeffwithfeedback} is nearly the CR energy density, the momentum diffusion coefficient will vanish at momenta $p>p_0$ if the energy density of the particles $e_{\rm CR}$ satisfies:
\begin{align}
 e_{\rm CR} &\gtrsim 3 c \( \frac{\rho^2 S p_0}{Z e B} \)^{1/3}
 \, ,
\end{align}
i.e. if the CR energy density is higher than typically 10~eV~cm$^{-3}$ for standard SB environments.
This estimate is consistent with that obtained by \citet{ptuskin2006} who concluded that the feedback of the particles on the turbulence spectrum of the ISM should be important below GeV energies. In young SBs, an energy density of 1-10 eV/cm$^3$ is already reached after the very first supernova explosion: the nonlinear CR feedback on the turbulence cascade must therefore be taken into account even for young clusters.

To obtain the turbulence spectrum, and thus the momentum diffusion coefficient, we solved the stationary version of Eq.~\ref{turbulence_cascade}. The timescale of the energy transfer at the scale $2 \upi/k$ is $\tau = \rho/(a \sqrt{W k^3})$. For a Kolmogorov spectrum, this timescale is always much smaller than the scattering time of the particles on the magnetic waves and the hypothesis of stationarity is well justified.


Together with stationarity, other simplifications have been done in the above derivation. We assumed that all modes of the turbulence followed a Kolmogorov-type cascade, whereas in realistic cosmic environments two cascades may coexist, one describing the dynamics of Alfvén waves and the other describing the fast magnetosonic modes \citep{ptuskin2003}.

\subsection{Spatial diffusion and escape}
When they propagate in the turbulent medium, not only particles are reaccelerated by stochastic scatterings on magnetic waves, but they also experience an effective spatial diffusion until they escape in the interstellar medium (ISM). Under the quasilinear hypothesis, the following approximate relation between the spatial and momentum diffusion coefficients is obtained for a Kolmogorov turbulence\footnote{The ``standard'' relation $D_x D_p = v_A^2 p^2/9$ is only valid in Bohm's regime \citep{thornbury2014}.}:
\begin{equation}\label{QLTDpDx}
    D_x D_p \approx 0.2 v_A^2 p^2
    \, .
\end{equation}
This relation, derived for a power law turbulence spectrum $W(k) \propto k^{-5/3}$, is also valid if small scales are damped, providing the largest scales do follow a Kolmogorov scaling.
In the test-particle regime, Eq.~\ref{diffusioncoeffwithfeedback} eventually provides the following estimate:
\begin{multline}
    D_x \approx 10^{28} \beta \(\frac{p}{m_p c}\)^{\frac{1}{3}} \( \frac{B}{10 ~ \text{\textmu G}} \)^{\frac{5}{3}} 
    \( \frac{\eta_T L_*}{10^{51} ~\text{erg/Myr}} \)^{-\frac{2}{3}}
    \\ \times 
    \( \frac{n}{0.01 ~\text{cm}^{-3}} \)^{-\frac{1}{3}} \( \frac{R_b}{100 ~\text{pc}}\)^{2} ~\text{cm}^2/\text{s}
    \, .
\end{multline}
In the case where the CR energy density is so high that the turbulence cascade terminates at a large scale such that $D_p$ vanishes (this happens when the quantity in parenthesis in Eq.~\ref{diffusioncoeffwithfeedback} vanishes), the above expression breaks down. In this case, low energy particles do not scatter on waves anymore, but rather follow the field lines of the background field. In a highly turbulent medium\footnote{In this case, the quasi-linear approximation used to compute the diffusion of high energy particles on the waves is understood as a local approximation.}, the background field has a coherence length of the order of $0.77/k_0$ \citep{casse2001}.  Low energy particles following the background field do experience a random walk with mean displacement $0.77/k_0$. From a macroscopic point of view, this gives rise to an effective spatial diffusion with diffusion coefficient about $0.3 v/k_0$. We conveniently interpolate the coefficient between the low and high energy regimes as:
\begin{equation}
    D_x \approx \( \frac{5 D_p}{v_A^2 p^2} + \frac{3 k_0}{v} \)^{-1}
    \, .
\end{equation}
The quasilinear relation given by Eq.~\ref{QLTDpDx} is retrieved whenever the particles scatter on magnetic waves, while in the absence of waves one gets $D_x \approx 0.3 v/k_0$.

In fact, the diffusion of CRs below the inertial turbulence cascade is a more complicated process than the simple picture presented above, in particular because the anisotropic streaming along the background field is expected to trigger the so-called streaming instability, such that the particles will scatter on the waves they produce. Accounting for the streaming instability requires the knowledge of the spatial gradient of the distribution function along the magnetic field lines, which is beyond the scope of this work. Furthermore, it is not straightforward to solve the turbulence cascade including the growth of the streaming instability. For simplicity, we assume that the background field is chaotic enough to efficiently isotropise the low energy particles, such that the streaming instability can be neglected. Moreover, although it is crucial to account for the feedback of the energetic particles on the hydromagnetic waves, the cascade is in fact very scarcely suppressed for the parameters we consider.

We eventually average the CR transport equation over the SB volume such that the spatial diffusion reduces to a leakage with characteristic escape time $\tau = R_b^2/(6 D_x)$. The complete averaged CR transport equation in the SB then reads:
\begin{equation}\label{SBequation}
    \d_t n =  - \frac{n}{\tau} + \frac{1}{p^2} \d_p \( p^2 D_p \d_p n \)
    + \frac{1}{p^2} \d_p \( p^2 \left[ \left. \frac{\dd p}{\dd t} \right|_{\rm Ad.} + \left. \frac{\dd p}{\dd t} \right|_{\rm Int.} \right] n \) + q_w \, ,
\end{equation}
where $n = V_{\rm SB} f(p)$ is the CR spectrum in the SB. The first loss rate in the right-hand side accounts for the adiabatic decompression of the CR spectrum in the expanding medium, $\left. \frac{\dd p}{\dd t} \right|_{\rm Ad.} = (\epsilon(p)/v) \dot{R_b}/R_b$ (e.g. \citealt{finke2012} or Appendix~B of \citealt{aharonian2004}). The second loss rate $\left. \frac{\dd p}{\dd t} \right|_{\rm Int.}$ encodes the losses due to the Coulomb interactions and to the production of pions \citep{mannheim1994}.

The computation of the CR dynamics in the SB follows the procedure used by \citet{ferrand2010}. In between supernova explosions, the spectrum evolves according to Eq.~\ref{SBequation} implemented within a forward Euler scheme (a Crank-Nicholson scheme can not be used since the momentum diffusion is nonlinear). Whenever a supernova explodes, the reacceleration of the particles is computed according to the formalism detailed in \citep{paper2}. All other physical processes are neglected during the supernova explosion.

The parameters used to model the fiducial SB considered in this work are summarised in Table~\ref{tab:parameters}.

\begin{table}
\centering
\begin{tabular}{lcc}
\hline
Initial number of massive ($M>8 M_\odot$) stars & $N_*$ & $100 - 1000$ \\
Superbubble radius normalised per star at 1 Myr & $R_0$ & 10 pc \\
ISM density around the SB & $n_{\rm ISM}$ & $100$ cm$^{-3}$ \\
Large scale magnetic field & $B_0$ & 10 \textmu G \\
Efficiency of turbulence generation & $\eta_T$ & $1 - 30$ \% \\
Turbulence relaxation time after SNR & $\tau_T$ & 1 Myr \\
Largest turbulent scale & $\lambda$ & 10 pc \\
Injection parameter of SNR shocks & $\xi$ & 3.5 \\
Wind acceleration efficiency & $\eta_w$ & 10 \% \\
Injection momentum at winds & $p_0$ & 5 MeV/c \\
Maximum momentum at winds & $p_w$ & 10$^5$ GeV/c \\
\hline
\end{tabular}
\caption{Exhaustive list of the parameters characterising the fiducial superbubble considered in this study.}
\label{tab:parameters}
\end{table}

\section{Results}\label{sec:results}
We now apply the model to the acceleration of protons in order to investigate their energetics and spectra in SB.

\subsection{Timescales}\label{sec:CRinSB:timescales}
\begin{figure}
\centering
  \includegraphics[width=\linewidth]{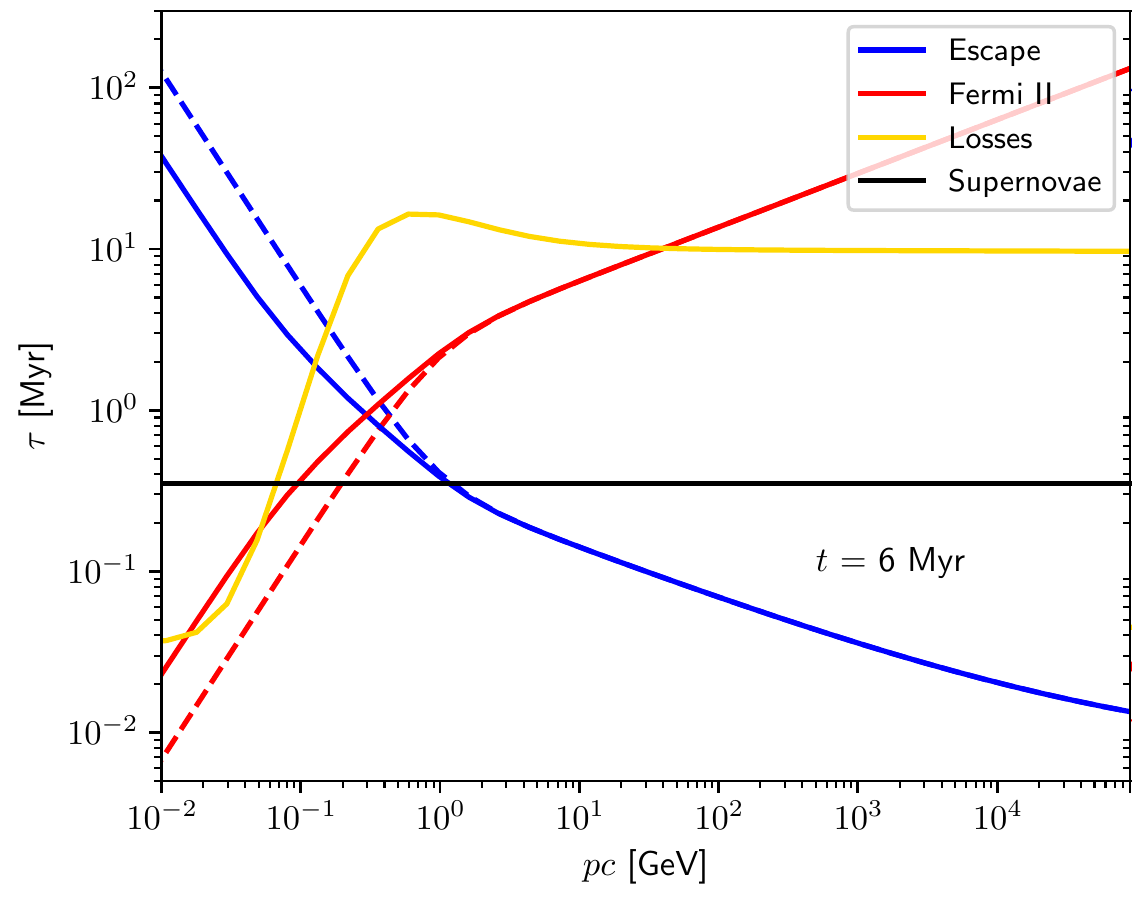}
\caption{Typical timescales of the fiducial superbubble (see the parameters in Table~\ref{tab:parameters}) for $N_* = 100$, $\eta_T = 30 \%$, and a cluster age of 6~Myr. The solid lines correspond to the nonlinear computation accounting for the CR feedback on the turbulence spectrum, to be compared with the dotted lines displaying the test-particle expectation. The horizontal black line indicates the typical time interval between two supernovae in a cluster of 100 massive stars (0.37~Myr).}
\label{fig:timescales}
\end{figure}
Figure~\ref{fig:timescales} displays the timescales of stochastic (Fermi II) acceleration $p^2/(2 D_p)$ as well as the escape time $\tau$ for the fiducial SB. At high energies ($p \gtrsim 10$~GeV/c), the standard Kolmogorov scalings are always retrieved. Below 10 GeV, the cascade can be damped if there are enough massive stars in the cluster. In this case, the stochastic acceleration is less efficient than expected in the test-particle limit, while the escape is faster. Sub-GeV energies are driven by the stochastic reacceleration while the escape dominates above GeV energies. If the number of massive stars were increased, supernovae would be more frequent and the horizontal black line in Figure 5 would be lowered. Supernovae would then be frequent enough to sustain the other processes in a broader energy range, between about 0.1 and 10~GeV, leading to a nearly stationary production of CR in this range. On the other hand, whenever the dominant process is faster than the SNR rate, the production of CR is expected to be intermittent. This is the case at low and high energies.

For completeness we added the timescale of the proton losses, including Coulomb collisions, protons-protons interactions, as well as adiabatic decompression. Proton losses are only relevant at the smallest energies. 

\subsection{Cosmic ray energetics in superbubbles}
\begin{figure*}
\centering
  \includegraphics[width=1\linewidth]{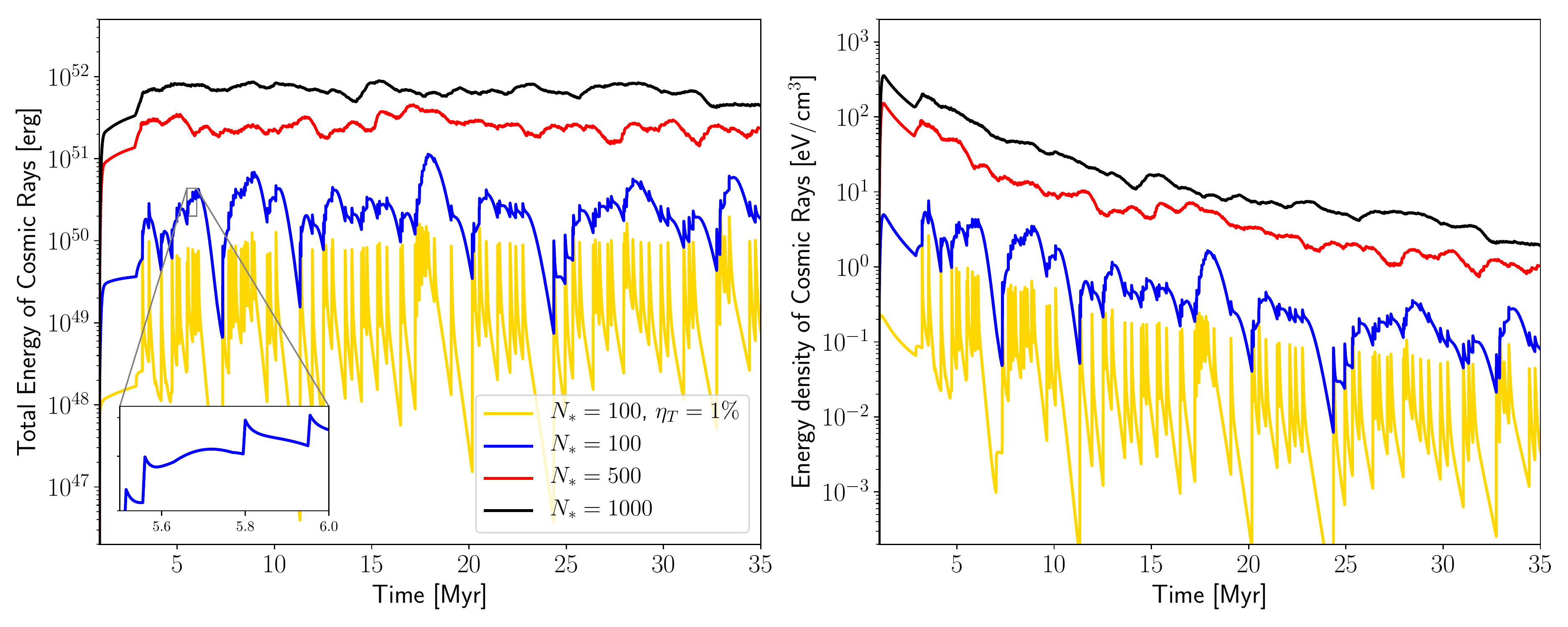}
\caption{Time evolution of the total CR energy (left) and CR energy density (right) inside the SB for clusters containing initially 100, 500 and 1000 stars. All the results are obtained with $\eta_T=30\%$ except those shown by the yellow curves for which $\eta_T=1\%$.}
\label{fig:evolution_energy}
\end{figure*}

Figure~\ref{fig:evolution_energy} displays the CR energetics in a fiducial SB described by the parameters given in Table~\ref{tab:parameters}. About 10\% of the mechanical power is transferred into CRs. For small clusters ($N_* < 100$), the CR production becomes intermittent, in particular if the turbulence is not efficiently generated, as shown by the yellow curve where we set $\eta_T = 1 \%$. In this case the particles are not confined and the SB is similar to a collection of individual SNR.

As shown by the zoomed-in window in the left panel of Figure~\ref{fig:evolution_energy}, CR acceleration proceeds in two steps. First, DSA at SNR injects nonthermal particles inside the SB. In the following few 10 kyr, the high energy particles rapidly escape. Meanwhile, low energy particles are reaccelerated in the magnetic turbulence. After around 100 kyr, this second order reacceleration compensates the escape and the total energy increases progressively. However, the process of stochastic reacceleration is only effective at low energies, such that when a substantial fraction of the particles has been reaccelerated, the escape dominates once again and the total energy decreases, until the next star explodes.

The right panel of Figure~\ref{fig:evolution_energy} displays the time evolution of the CR energy density inside the SB. Densities of the order of 1 to 10 eV/cm$^3$ are expected in rather small clusters ($N_* < 100$), while larger clusters gathering hundreds of stars may contain a CR energy density higher than 100 eV/cm$^3$.

Our computation also shows that even when accounting for nonlinear effects, CRs are mainly injected by SNRs (e.g. $95$\% of the particles come from SNRs for $\eta_w=0.1$, or $78$\% for $\eta_w=0.5$). The fraction is slightly smaller for more massive clusters (e.g. 64\% for a cluster of 500 stars with $\eta_w=0.5$) because the nonlinear effects are stronger. Yet, the supernovae are dominant in most cases. This result matches the requirement needed to reproduce the composition of galactic CRs \citep{tatischeff2021}.

In fact, the energy transfer between the winds and the CRs proceeds mainly indirectly through the stochastic reacceleration in turbulence. As the low- and intermediate-energy particles are confined in the SB, this process is very efficient. Nearly all the energy injected from the stars to the turbulence is transferred to CRs. This could explain the high gamma-ray fluxes detected in e.g. G25.18+0.26 \citep{katsuta2017}, without requiring an unrealistically small diffusion coefficient. This statement has to be confirmed by further studies including the inhomogeneity of the turbulence and distribution function. Indeed, if the stars are gathered at the centre of the cluster, they are expected to excite Alfvén waves only in a small region of space. On the other hand, the CR density is expected to be higher close to the centre of the bubble, such that most of the particles could still be efficiently reaccelerated.

Because the main source of matter inside the SB is the evaporation of interstellar material at the shell interface, most of the material accelerated in the SB by SNRs is expected to be of galactic composition. A discrepancy is however expected because of the efficient injection of WR-enriched material in the early phase of the cluster life. Although the fraction of CRs accelerated at WTS is only about 5-10\% of that accelerated at SNRs, this may be enough to account for the observed overabundance of $^{22}$Ne over $^{20}$Ne in the galactic CRs, in particular if very massive stars collapse without exploding \citep{kalyashova2019,gupta2020,tatischeff2021}.

\subsection{Intermittency}
\begin{figure*}
\centering
  \includegraphics[width=1.\linewidth]{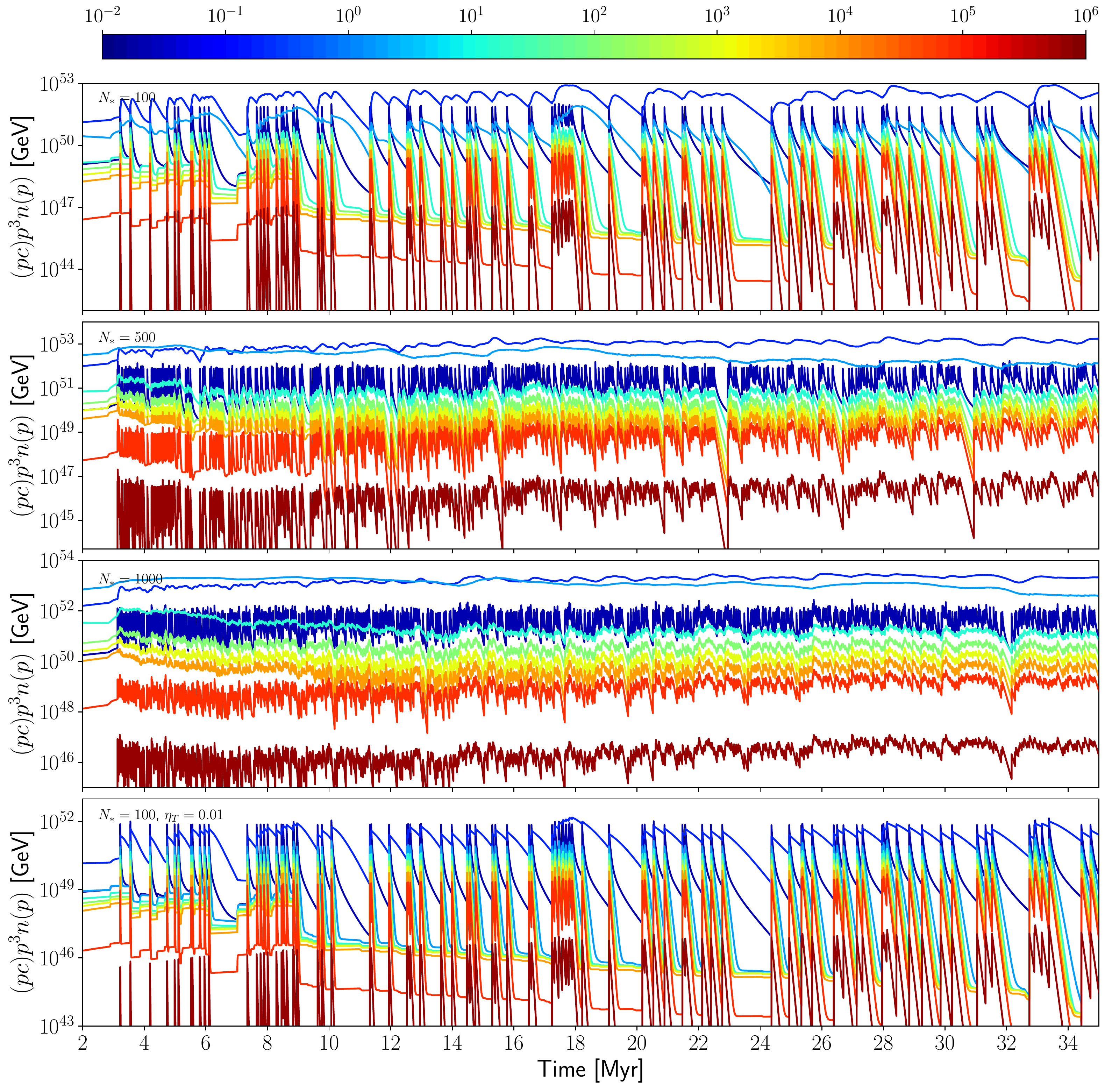}
\caption{Time evolution of the particle distribution function. The parameters of the SB are given in Table~\ref{tab:parameters}. Top panel: 100 massive stars. Second panel: 500 massive stars. Third panel: 1000 massive stars. Bottom panel: 100 massive stars, with $\eta_T = 1 \%$ instead of 30\%. Colours correspond to different momenta (see the colour scale in units of GeV/c at the top). Dark blue: 10~MeV/c. Blue: 0.13~GeV/c. Light blue: 1.6~GeV/c. Turquoise blue: 20~GeV/c. Green: 240~GeV/c. Yellow: 2.9~TeV/c. Orange: 35~TeV/c. Red: 0.43 PeV/c. Dark red: 5.3~PeV/c.}
\label{fig:intermittency}
\end{figure*}
The time evolution of the particle distribution function is shown in Figure~\ref{fig:intermittency} for various initial numbers of massive stars. The low energy bands ($p\lesssim 0.1$~GeV/c) are always intermittent because low energy particles are stochastically reaccelerated in the turbulence as soon as they are injected into the SB. Intermediate energies are less intermittent because the stochastic acceleration is less efficient and the escape is slow. In particular, for standard clusters containing hundreds of massive stars, the GeV band is nearly stationary, except if the turbulence level is low (bottom panel of Figure~\ref{fig:intermittency}), in which case even these rather low energies are not confined and escape between supernova explosions. High energies are very intermittent because the escape is faster than the interval between two supernova explosions. Three regimes of CR production should therefore be distinguished. Whenever the stochastic acceleration time is lower than the average time interval between two supernovae, i.e. $p^2/2 D_p(p) < \Delta t_{\rm SN}$, the CR production is expected to be intermittent. This regime extends up to the momentum $p_1$ defined by $p_1^2/2 D_p(p_1) = \Delta t_{\rm SN}$. In the test-particle approximation, we obtain:
\begin{multline}\label{CR_in_SB:p1}
    \beta(p_1)^3 p_1 = 4 {~\rm MeV/c} \( \frac{N_*}{100} \)^{-3} \( \frac{\eta_T L_*}{10^{51} {~\rm erg/Myr}} \)^2 \( \frac{\mathcal{T}_{\rm SB}}{35 {~\rm Myr}} \)^{3}
    \\ \times
    \( \frac{n}{0.01 ~{\rm cm}^{-3}} \)^2 \( \frac{B}{10 {~\text{\textmu G}}} \) \( \frac{R_b}{50 ~{\rm pc}} \)^{-6}
    \, .
\end{multline}
Because the damping of the turbulence by non-thermal particles is generally moderate, the test-particle approximation is expected to give reliable estimates on average, even though the nonlinear timescale of diffusion may decrease considerably right after a supernova explosion.

\begin{figure*}
\centering
  \includegraphics[width=0.8\linewidth]{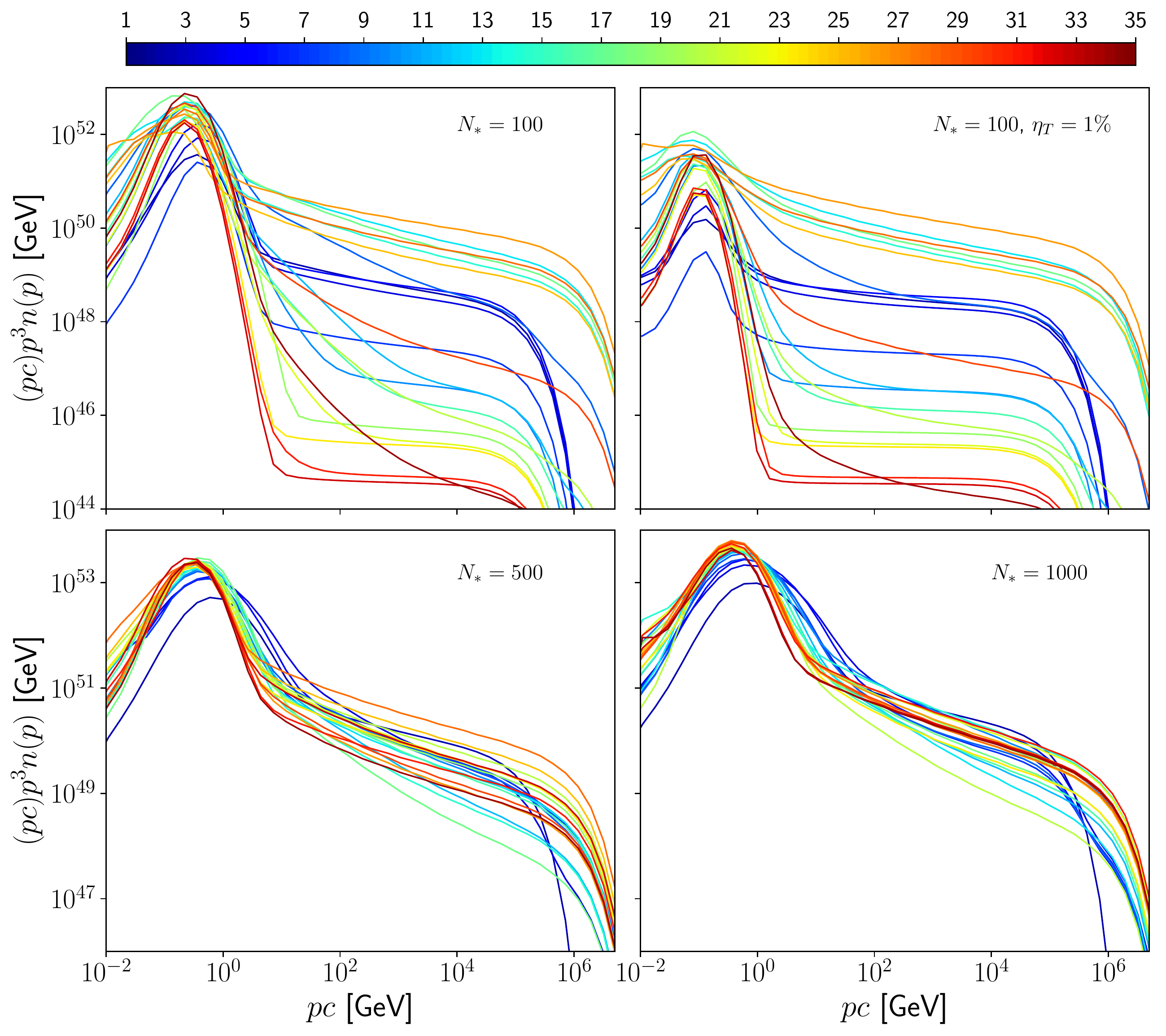}
\caption{Instantaneous SB spectra. Each curve corresponds to a different time represented by the colour scale in units of Myr at the top. The SB parameters are given in Table~\ref{tab:parameters}, with varying number of stars. Top left panel: $N_* = 100$. Top right panel: $N_* = 100$ with $\eta_T = 1\%$. Bottom left panel: $N_* = 500$. Bottom right panel: $N_* = 1000$.}
\label{fig:spectra}
\end{figure*}

Then, if both the stochastic acceleration time as well as the escape time are larger than the average time interval between two supernovae, i.e. $p^2/2 D_p(p), R_s^2/6 D_x(p) < \Delta t_{\rm SN}$, the CR production is expected to be nearly stationary. This regime extends up to the momentum $p_2$ defined by $R_s^2/6 D_x(p_2) = \Delta t_{\rm SN}$, which translates into:
\begin{multline}\label{CR_in_SB:p2}
    p_2 = 0.09 {~\rm GeV/c} \( \frac{N_*}{100} \)^3 \( \frac{\eta_T L_*}{10^{51} {~\rm erg/Myr}} \)^2 \( \frac{\mathcal{T}_{\rm SB}}{35 {~\rm Myr}} \)^{-3}
    \\ \times
    \( \frac{n}{0.01 ~{\rm cm}^{-3}} \) 
     \( \frac{B}{10 {~\text{\textmu G}}} \)^{-5} \, .
\end{multline}
Finally, for $p>p_2$, the escape dominates and the cosmic ray production is expected to be intermittent again. The three regimes can be identified in Figure~\ref{fig:timescales}, comparing the Fermi II acceleration time (red curve), the escape time (blue curve) and the time interval between supernovae (black line). If the black line is below the other timescales, supernovae can sustain the reacceleration and escape. In other cases, intermittent reacceleration or escape is expected.

It may happen that $p_2<p_1$. In this case, there are not enough stars to sustain a stationary CR production against the escape, even at intermediate energies. This translates into an intermittent CR production at all energies, as it is the case for the cluster of 100 stars with inefficient turbulence generation ($\eta_T=1\%$, see bottom panel of Figure~\ref{fig:intermittency}).

For our fiducial clusters containing respectively 100, 500 and 1000 massive stars, Equations~\ref{CR_in_SB:p1} and \ref{CR_in_SB:p2} provide respectively $p_1 \approx 0.1$,~0.04,~0.01~GeV and $p_2 \approx 1$~GeV,~3~TeV,~1~PeV, which is consistent with the curves shown in Figure~\ref{fig:intermittency}. In these cases, the test-particle estimates are proved to be reliable indicators within one order of magnitude.

Eventually, one notices in the top and bottom panels of Figure~\ref{fig:intermittency} that the winds provide a stationary background below the intermittent emission due to supernova explosions. Although the wind contribution is subdominant below about 10~GeV; it is non negligible at higher energies, where the escape is efficient.

\subsection{Spectra}
Examples of spectra resulting from our nonlinear computation are shown in Figure~\ref{fig:spectra}. The overall shape agrees qualitatively with the conclusions drawn by \citet{ferrand2010}, that is, a hard component at low momenta competes with a steep component at high momenta. We denote $p_*$ the momentum of the transition. As we expressed the turbulence spectrum as function of the mechanical energy of the stars rather than as function of the background magnetic field, our expression for $p_*$ differs from that given in \citet{ferrand2010} even in the test-particle regime:
\begin{multline}\label{TPestimatepstar} 
    \beta(p_*)^3 p_* \approx 0.1 m_p c \( \frac{B}{10 ~ \text{\textmu G}} \)^{-2}
 \( \frac{\eta_T L_*}{10^{51} ~\text{erg/Myr}} \)^{2} \\ \times 
 \( \frac{n}{0.01 ~\text{cm}^{-3}} \)^{-1/2} \( \frac{R_b}{100 ~\text{pc}}\)^{-3}
 \, .
\end{multline}
One notices that $p_*$ decreases as the SB expands, which is a consequence of the dilution of the mechanical energy of the stars in an increasing volume. In the fiducial cluster of 100 massive stars considered in this work, $p_*$ is found around a few to tens of GeV. In typical clusters of several hundreds of stars, $p_*$ can hardly be higher than a few tens of GeV, even considering unrealistic turbulence generation efficiencies, because in this case nonlinearities are expected to regulate the stochastic acceleration process by damping the waves.

The shapes of the SB spectra are overall well understood by the timescales of the competing reacceleration and escape processes.
One may wonder why the high energy ``universal'' asymptotic solution of successive shock reacceleration computed in \citet{paper2} is never retrieved, in particular for the cluster of 1000 massive stars, where the average time interval between supernovae is about 30~kyr, that is, much smaller than the escape time at GeV energies. This is because we implicitly assumed, by considering the average CR spectrum $n(p)$ instead of the distribution function $f(x,p)$, that the (re)accelerated CRs instantaneously homogenise inside the SB. Under these circumstances, preaccelerated particles have small chances to be reaccelerated by the next SNR shock, which only spans a few percent of the SB volume. Such a situation corresponds to a loose cluster, with large distances between the stars. The model will be refined in Section~\ref{sec:6:compactclusters} in order to include the case of a compact cluster.

Finally, although the contribution from the winds is globally subdominant, it yet provides a non-negligible nearly stationary steep component (see Figure~\ref{fig:spectra}).

\begin{figure}
\centering
  \includegraphics[width=\linewidth]{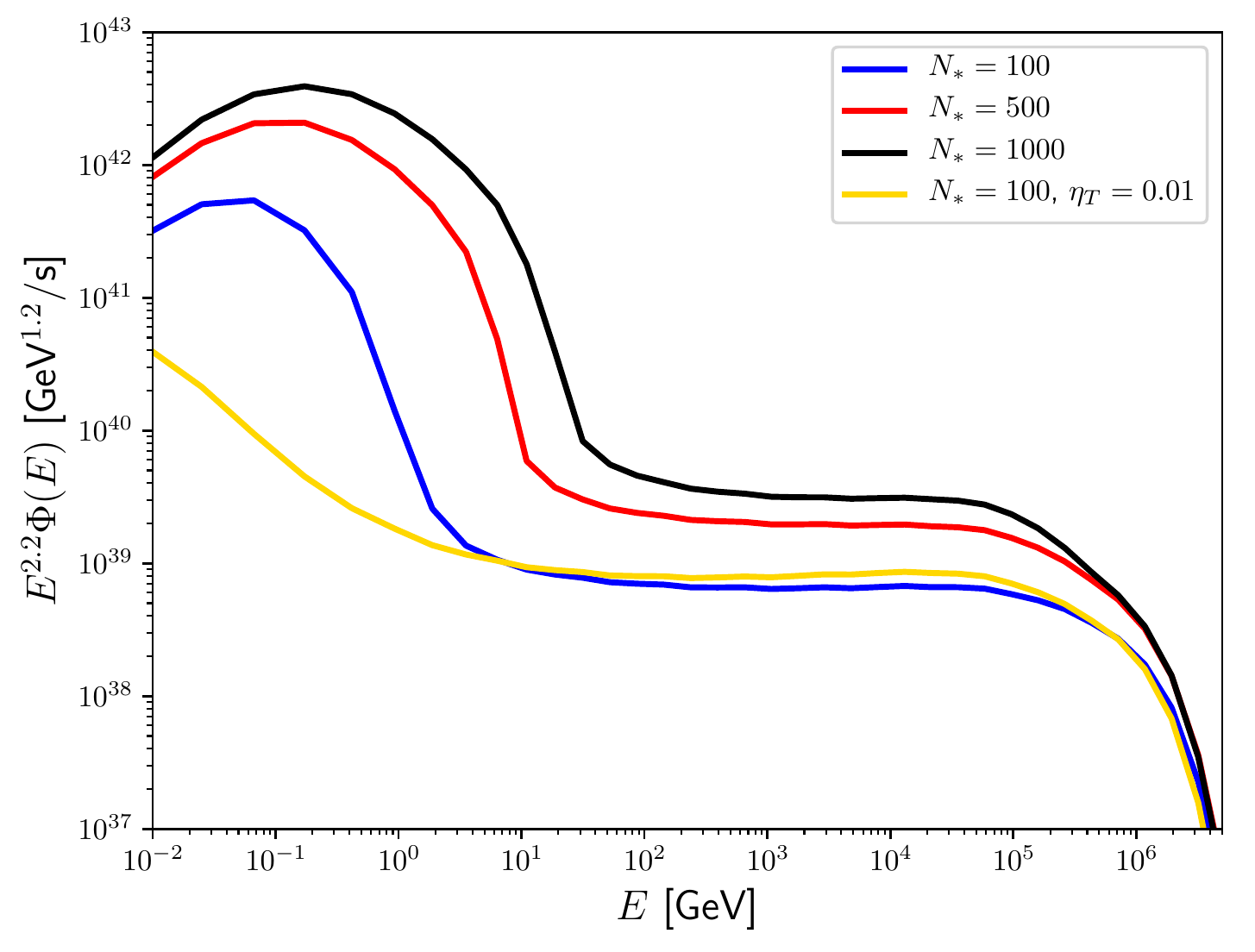}
\caption{Average CR flux escaping the SB.}
\label{fig:escapespectra}
\end{figure}

When integrated over the lifetime of the cluster, the spectra display an enlarged ``Fermi II bump'' and then a steep high energy tail above 10-100 GeV, as shown in Figure~\ref{fig:escapespectra}, where the spectra have been corrected as $\Phi(E) = 4 \upi p^2 n(p)/(v(p)\tau(p))$. This corresponds to the CR flux escaping from the cluster in the ISM, except above 1~PeV where we disregarded the very high energy flux escaping upstream of SNRs. Interestingly, the high energy flux is a power law scaling as $\Phi(E) \propto E^{-2.2}$, which would reproduce the $E^{-2.7}$ scaling measured around the Earth if one assumes that the CR diffusion coefficient scales as $D(E) \propto E^{0.5}$ in the ISM, as expected in the Iroshnikov-Kraichnan regime of turbulence. Eventually, the transition between the maximum energy achievable in WTS (assumed to be about 100~TeV) and that achieved at SNRs (about 1~PeV), produces a spectral break between 0.1 and 1~PeV which has a certain resemblance to the knee observed in the galactic CR spectrum.

\section{Two-zone model}\label{sec:5}
In this section we show that it is possible to refine the modelling in order to compute supernova reacceleration in compact clusters in a more realistic way as well as to consider the effect of the SB supershell.

\subsection{Diffusion in a two-zone model}
\begin{figure}
\centering
  \begin{minipage}[b]{0.6\linewidth}
   \centering
   \includegraphics[width=\linewidth]{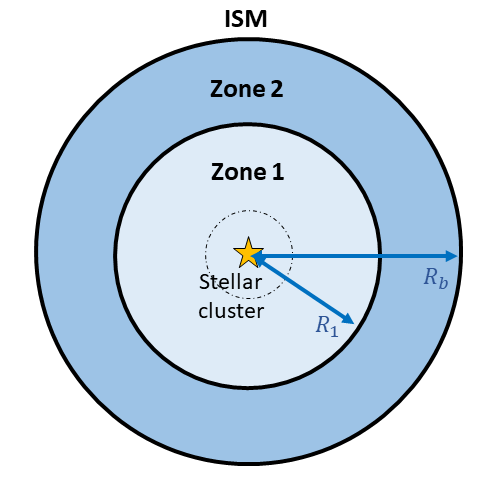}
  \end{minipage}
\hfill
  \begin{minipage}[b]{0.49\linewidth}
   \centering
   \includegraphics[width=\linewidth]{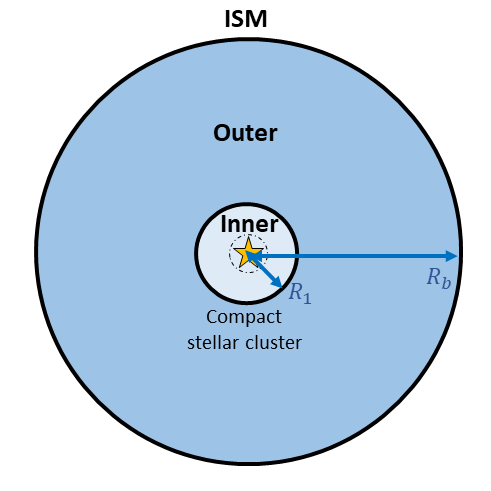}    
  \end{minipage}
\hfill
   \begin{minipage}[b]{0.49\linewidth}
   \centering
   \includegraphics[width=\linewidth]{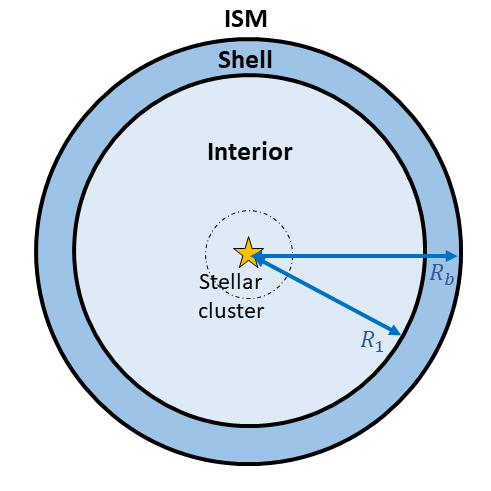}    
  \end{minipage}
  \caption{Top: Sketch of the two-zone model. Bottom left: Limit case of a compact cluster, where zone 1 is the region of supernova reacceleration. Bottom right: Limit case of a supershell.}
  \label{fig:sketch2zones}
\end{figure}

Let us split the interior of the spherical SB into two zones, as schemed in Figure~\ref{fig:sketch2zones}. The top panel presents the general setup while the two other panels display two limit cases of physical interest, which will be discussed below. In this section we stick to a general formalism, assuming that the inner region has a radius $R_1$ and the outer region has a radius $R_b$, which is the SB radius. The boundaries of both areas are therefore separated by a length equal to $R_b - R_1$. 
The two regions are characterised by diffusion coefficients $D_1$ and $D_2$, respectively. We  denote $f_1$ and $f_2$ the average distributions of particles in both regions such that the corresponding energy densities are $n_1 = V_1 f_1$ and $n_2 = V_2 f_2$, where $V_1 = 4/3 \upi R_1^3$ and $V_2 = V_{\rm SB} - V_1$ with $V_{\rm SB}$ the SB volume.

\begin{figure*}
\centering
  \includegraphics[width=\linewidth]{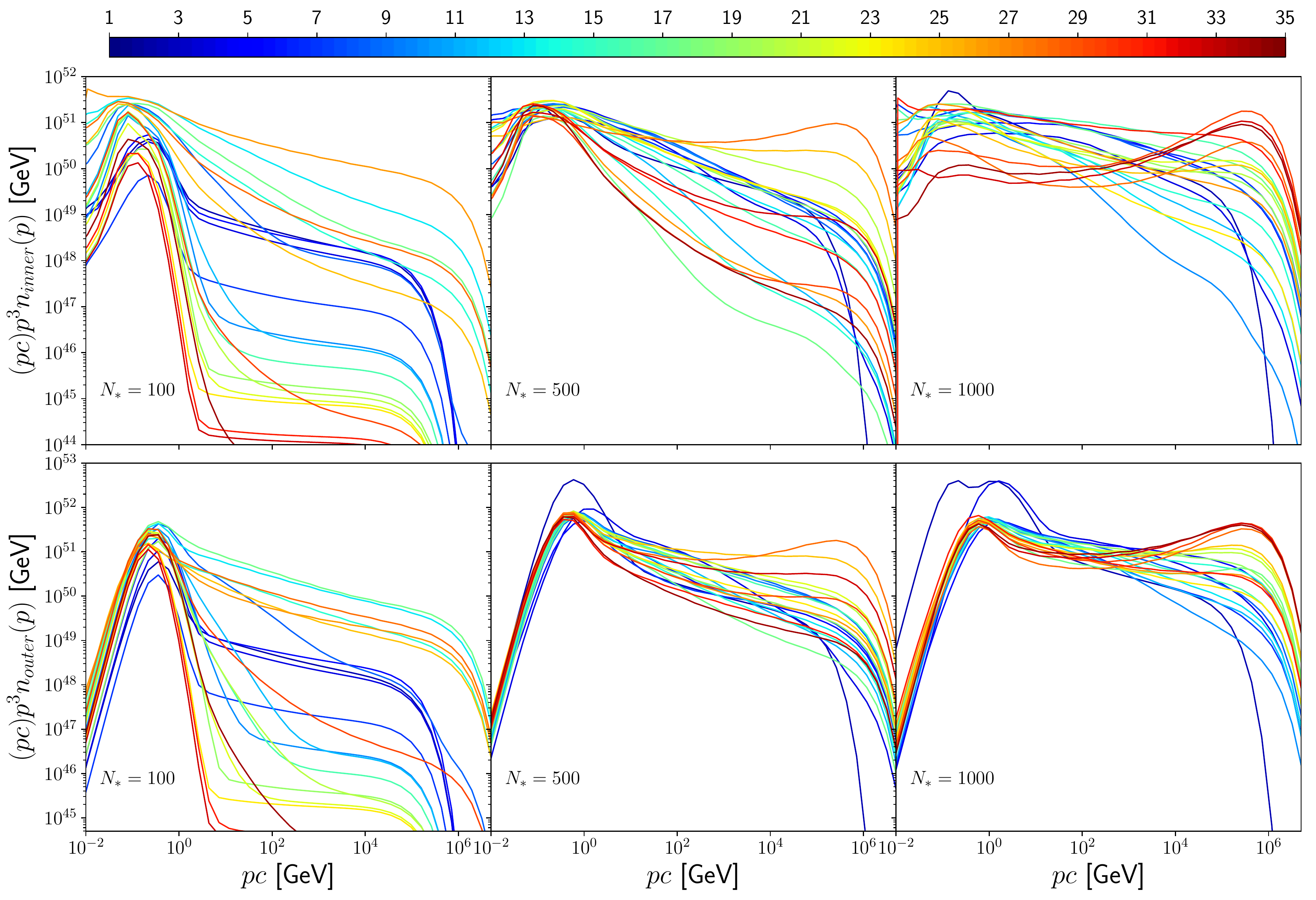}
\caption{Average spectra around a compact cluster modelled as a two-zone model (inner/outer). Each curve corresponds to a different time represented by the colour scale in units of Myr at the top. The number of massive stars is assumed to be 100 (left), 500 (middle) and 1000 (right). The top panels show the spectrum in the inner region spanned by SNRs while the bottom panels show the spectrum in the outer region, which fills most of the SB volume.}
\label{fig:spectrareacczone}
\end{figure*}

The CR transport in the SB is then described as a two-zone diffusion-losses model. The transport equation reads:
\begin{equation}\label{CR_in_SB:twozones}
\d_t f_i = \nabla \cdot \( D_i \cdot \nabla f_i  \) + \mathcal{D}_i[f_i] + \mathcal{L}_i [f_i] + q_i \, ,
\end{equation}
where $i = 1$ (resp. $i=2$) in the inner (resp. outer) region, $\mathcal{D}_i[f_i] = (1/p^2) \d_p \left( p^2 D_{p,i} \d_p f_i \right)$ is the stochastic reacceleration operator, with $D_{p,i}$ the momentum diffusion coefficient and $\mathcal{L}_i$ an operator encoding the various loss processes in both zones, and $q_i$ encodes the background wind injection in zone 1 ($q_1 = q_w$, with $q_w$ defined in Eq.~\ref{qwindinjection}) and is zero in zone 2. Instead of solving directly Eq.~\ref{CR_in_SB:twozones}, we average the distribution functions $f_i$. Denoting $\tau_{12}$ the typical time it takes for particles to diffuse from zone 1 to zone 2 (and conversely), and $\tau_2$ the escape time from zone 2 to the ISM, we obtain the following system of differential equations:

\begin{equation}\label{paper3:TEtwozones}
\left\lbrace
\begin{array}{lll}
\d_t n_1 & = & \frac{n_2-n_1}{\tau_{12}} + \mathcal{D}_1[n_1] + \mathcal{L}_1[n_1] + q_w \, , \\
\d_t n_2 & = & \frac{n_1-n_2}{\tau_{12}} - \frac{n_2}{\tau_2} + \mathcal{D}_2[n_2] + \mathcal{L}_2[n_2] + \phi_{\rm ISM} \, , \\
\end{array}\right.
\end{equation}
where $\phi_{\rm ISM}$ is the flux of interstellar particles entering the outer region.

To estimate the typical residence time of particles in between zone 1 and zone 2, we compute the diffusion time between the centre of the bubble ($r=0$) and the middle of the shell ($r=(R_1 + R_b)/2$). To do so, we follow the approach presented in e.g. \citet{drury1983}. The solution of the stationary diffusion equation in spherical symmetry, $\d_r (r^2 D \d_r f) = r^2 Q \delta(r)$, with a constant injection of particles $Q$ at $r = 0$, reads:
\begin{align}
\left\lbrace
\begin{array}{l}
r < R_1 ~~\text{(zone 1)} \implies f = \frac{Q}{4 \upi D_1} \( \frac{1}{R_1} - \frac{1}{r} \) + \frac{Q}{4 \upi D_2} \( \frac{2}{R_1 + R_b} - \frac{1}{R_1} \)
\, ,
\\
r > R_1 ~~\text{(zone 2)} \implies f = \frac{Q}{4 \upi D_2} \( \frac{2}{R_1 + R_b} - \frac{1}{r} \) 
\, . \\
\end{array}\right.
\end{align}
Integrating this solution from $0$ to $R_1$ and then from $R_1$ to $(R_1 + R_b)/2$, we obtain $N$, the total number of particles in between both regions:
\begin{equation}
N = \frac{Q R_1^2}{6 D_1} \( 1 + \frac{\rho_b}{\delta} + \frac{\rho_b^2}{4 \delta} \) \, ,
\end{equation}
where we set $\rho_b \equiv (R_b - R_1)/R_1$ and $\delta \equiv D_2/D_1$. The typical residence time is therefore identified as:
\begin{equation}\label{timescaletransfertwozones}
\tau_{12} = \frac{R_1^2}{6 D_1} \( 1 + \frac{\rho_b}{\delta} + \frac{\rho_b^2}{4 \delta} \) \, ,
\end{equation}
which can also be interpreted as the time it takes for a particle to be transferred from one zone to the other.

Eventually, we compute the escape time from zone 2 to the interstellar medium by integrating Eq.~\ref{CR_in_SB:twozones} over the volume of zone~2, assuming that the spatial diffusion coefficient in the ISM is much larger than that in zone~2. This provides, by virtue of the divergence theorem:
\begin{equation}
\d_t n_2 = \oiint \( D_2 \cdot \nabla f \) \cdot \dd \v{S} + \mathcal{D}_2 \left[ n_2 \right] + \mathcal{L}_2 \left[ n_2 \right] \, .
\end{equation}
Then we estimate the gradient of particles across the surface $S_b = 4 \upi R_b^2$ by means of a linear extrapolation as $\nabla f_2 \sim 2 \alpha (f_{\rm ISM} - n_2/V_2)/(R_b - R_1)$, with $\alpha$ a geometrical factor. The surface integral becomes trivial and we eventually get:
\begin{multline}\label{paper3:integte}
\d_t n_2 = \frac{6 \alpha D_2 R_b^2 (V_2 f_{\rm ISM} - n_2)}{(R_b^2 + R_b R_1 + R_1^2)(R_b-R_1)^2} + \mathcal{D}_2 \left[ n_2 \right] + \mathcal{L}_2 \left[ n_2 \right] 
\\
- \iint_{S_1} \( D_2 \cdot \nabla f \) \cdot \dd \v{S} \, ,
\end{multline}
where $f_{\rm ISM}$ is the diffuse CR distribution function in the ISM and $S_1$ the surface of the interface between the two zones. The surface integral over the inner surface $S_1$ encodes the transfer of particles from zone 1 to zone 2 which has been already computed.
From Eq.~\ref{paper3:integte} we identify the escape time from zone 2 to the ISM as well as the flux of interstellar particles entering the SB:
\begin{align}\label{tau2shell}
\tau_2 &= \frac{(R_b^2 + R_b R_1 + R_1^2)(R_b-R_1)^2}{6 \alpha D_2 R_b^2} \, ,
\\
\phi_{\rm ISM} &= \frac{4 \upi}{3 \tau_2} \( R_b^3 - R_1^3 \) f_{\rm ISM} \, .
\end{align}
In the limit case $R_1 \ll R_b$, one should set $\alpha = 1$ to get $\tau_2 = R_b^2/(6 D_2)$ which is the diffusion time in spherical symmetry. On the other hand, if $R_1 \approx R_b$, one should set $\alpha = 4$ in order to get $\tau_2 = ((R_b-R_1)/2)^2/(2 D_2) $, which is the diffusion time in one dimension.

Now that all timescales have been determined, Eq.~\ref{paper3:TEtwozones} can be written in matrix form and solved iteratively. Such two-zone model can be used to compute for instance the acceleration and transport of CRs in a SB where turbulence is only efficiently generated around the stellar cluster, or to account for the residence time of particles in the inner region spanned by successive SNRs, or to model the shielding of the interstellar CRs by the dense shell, or last but no least, to probe the modulation of the CR spectra induced by the supershell with associate non-thermal emission of photons.

\subsection{Cosmic ray reacceleration in compact clusters}\label{sec:6:compactclusters}

As mentioned in the end of Section~\ref{sec:results}, the one-zone model does not properly account for CR reacceleration in compact clusters, for it is implicitly assumed that the accelerated particles homogenise instantaneously in the SB after the passage of a remnant. The two-zone model can be used to describe the reacceleration process more accurately. Indeed, the inner region can be identified with the volume spanned by SNRs $\mathcal{V}_{\rm SNR}$\footnote{Because this volume depends on the density of the SB interior, it is time-dependent and the radius of the inner region is defined dynamically.}. The SNR radius is typically 10\% the SB radius such that this situation is close to the limit case $R_1 \ll R_b$ sketched in the bottom left panel of Figure~\ref{fig:sketch2zones}. For simplicity we assume that the turbulence is homogeneous in both zones, although in reality it should probably be stronger in the inner zone close to the energy deposition region. In the case of a compact cluster, a collective wind termination shock (WTS) is expected to form at a distance of a few tens of pc around the cluster. \citet{morlino2021} showed that a standard $p^{-4}$ spectrum is expected up to a maximum energy of hundreds of TeV, at most PeV. Therefore, Eq.~\ref{qwindinjection} satisfactorily reproduces the injection of particles at a collective WTS. In this section, we stick to a maximum energy of 100~TeV achieved at the WTS, as a conservative estimate. Eventually, we neglect the flux of interstellar particles leaking inside the SB ($\phi_{\rm ISM} = 0$). This assumption is justified for clusters hosting several hundreds of massive stars as the energy density of the confined particles is expected to be much larger than the interstellar CR energy density. Besides, low energy ISM particles will be depleted by the ionisation losses in the SB shell, while the steep high energy tail will always be negligible. For small clusters, the ISM may only provide a noticeable contribution to the SB content around GeV energies. Although it is straightforward to include this contribution in our model, we chose to discard it in order to make the interpretation of the spectra easier. In particular, this allows to directly compare the output of the two-zone model with the results of Section~\ref{sec:results}.

Figure~\ref{fig:spectrareacczone} displays the spectra obtained in this refined model. The first thing to notice is that the overall normalisation is about one to two orders of magnitude below that obtained using a one-zone model (Figure~\ref{fig:spectra}). The energy transfer between the SNR and the CRs is much less efficient because of the backreaction of the CRs onto the shocks in the inner region. Indeed, preaccelerated CRs do not have time to entirely dilute inside the SB between two supernova explosions. The particle energy density in the inner region is high, which slows down the shock precursor and makes the acceleration of low-energy particles less efficient \citep[see][]{paper2}. On the other hand, if the massive stars are numerous enough, the time interval between supernovae can be smaller than the escape time at intermediate or even high energies and the particles are efficiently reaccelerated without leaving the region. As expected from the ``benchmark'' asymptotic solution discussed in \citet{paper2}, the spectrum resulting from the reacceleration by successive shocks is concave, with a high energy slope close to 3.5 in the case where the escape of the particles is completely neglected. As shown in the middle panel of Figure~\ref{fig:spectrareacczone}, concave spectra are indeed intermittently retrieved in a cluster containing initially 500 massive stars. For a cluster of 1000 stars, as shown in the right panel of Figure~\ref{fig:spectrareacczone}, the reacceleration is very efficient and the benchmark asymptotic spectrum may be achieved up to the highest energies. Then the concave spectrum realised in the inner region is transferred in the outer region, where low energy bands experience a stochastic reacceleration. The particles will eventually interact with the shell and concave gamma-ray spectra should be expected, which could be a typical signature of massive compact clusters.

Interestingly, hard gamma-ray spectra with slight concavities were recently observed in the G25 region \citep{katsuta2017}. Although the nature of the source is yet unclear, it was suspected that a star forming region could be responsible for this extended emission. Such spectra are indeed possible signatures of shock reacceleration in a SB surrounding the hypothetical G25.18+0.26 young stellar association.

\begin{figure*}
\centering
  \includegraphics[width=0.9\linewidth]{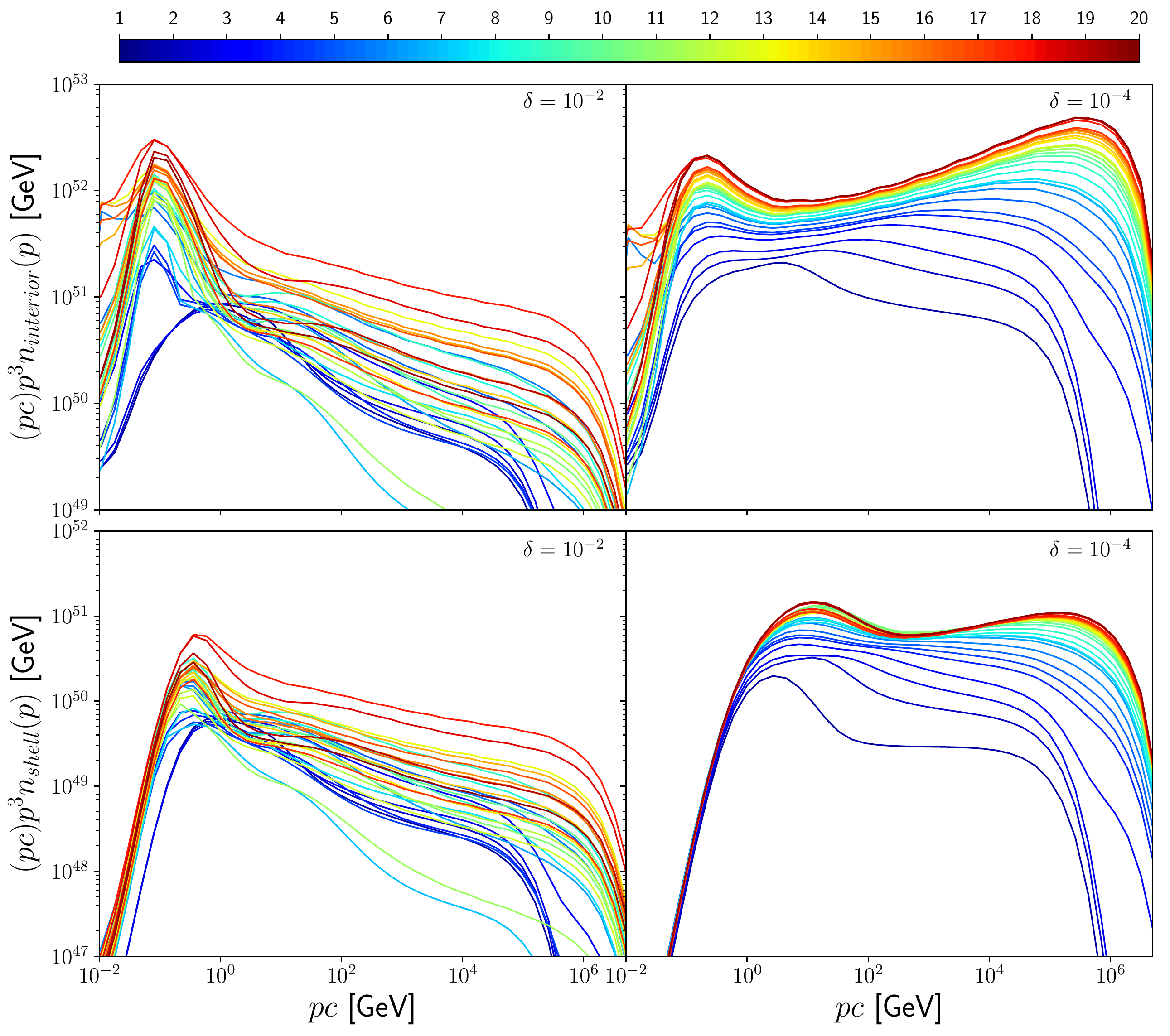}
\caption{Average spectra in a two-zone (top panels: interiors, bottom panels: shells) model with two different values of the shell confinement coefficient $\rho_b/\delta$ = 10 (left panels), 1000 (right panels). Each curve corresponds to a different time represented by the colour scale in units of Myr at the top.}
\label{fig:spectrashell}
\end{figure*}

\subsection{Modelling the supershell}
The two-zone model can also be used to account for the effect of the dense supershell in which the interstellar matter and magnetic field swept-up by the SB forward shock accumulate. The interstellar magnetic field is compressed and strengthened in the direction tangential to the shell. In order to escape in the interstellar medium, particles must diffuse mainly perpendicularly to this strong background field. Deriving the expression for a perpendicular diffusion coefficient from first principles is far from being trivial. On the other hand, heuristic arguments as well as numerical simulations suggest that $D_\perp \approx (\delta B/B)^{\alpha_B} D_\parallel$, with $\alpha_B \approx 4$ \citep[e.g.][]{casse2001,mertsch2020}. Therefore, the diffusion coefficient in the shell could well be orders of magnitude below that in the superbubble interior, which would enhance the confinement of the particles. This can be accounted for in the two-zone model as the parameter $\delta$ in Eq.~\ref{timescaletransfertwozones} is precisely the ratio of the diffusion coefficients. Eventually we need to specify the shell density and thickness. For simplicity we assume that the shell thickness is 10\% of the SB radius at any time, that is, $\rho_b = 0.1$. This parameter is not well constrained. Observations and numerical simulations provide estimates ranging from a few percent \citep[e.g.][]{joubaud2019} to tens of percent \citep[e.g.][]{krause2013} and even up to 50\% \citep[e.g.][]{yadav2017}. The hydrogen number density in the shell $n_H$ follows by assuming that all the mass swept-up by the SB forward shock has accumulated in the shell, which provides $n_H = n_{\rm ISM}(1+\rho_b)^3 /((1+\rho_b)^3-1)$. In the following we assume $n_{\rm ISM} = 10$~cm$^{-3}$.

Eq.~\ref{timescaletransfertwozones} predicts that the confinement of the particles will be enhanced when $\rho_b/\delta > 1$. Indeed the left panels of Figure~\ref{fig:spectrashell} show that for $\delta = 0.01$ ($\rho_b/\delta = 10$), the emission in a SB hosting 100 massive stars is much less intermittent, which results in overall smoother and steeper spectra because the flat wind component is always subdominant. For a high ``shell confinement coefficient'' $\rho_b/\delta = 10^3$ (right panels of Figure~\ref{fig:spectrashell}), the particles are strongly confined. In this case, the CR energy density in the shell is so high that the turbulence cascade terminates close to the injection scale: the shell is not turbulent. This demonstrates that low values of the parameter $\delta$ (that is, strong confinements) are not unrealistic, although a self-consistent model computing $\delta$ from first principles and accurately accounting for the backreaction of the particles on the perpendicular transport is still missing.

Whenever the shell efficiently enhances the confinement, the particles reaccelerated at SNRs stay inside the SB in such a way that the interior is characterised by a homogeneous concave spectrum which converges toward the asymptotic solution discussed in \citet{paper2}. Eventually, one notices in the bottom panels of Figure~\ref{fig:spectrashell} that the spectra are cut off below 1~GeV, which is due to the enhanced energy losses suffered by low-energy particles within the dense shell.

\section{Superbubble contribution to galactic cosmic rays}\label{sec:6}
\begin{figure}
\centering
  \includegraphics[width=\linewidth]{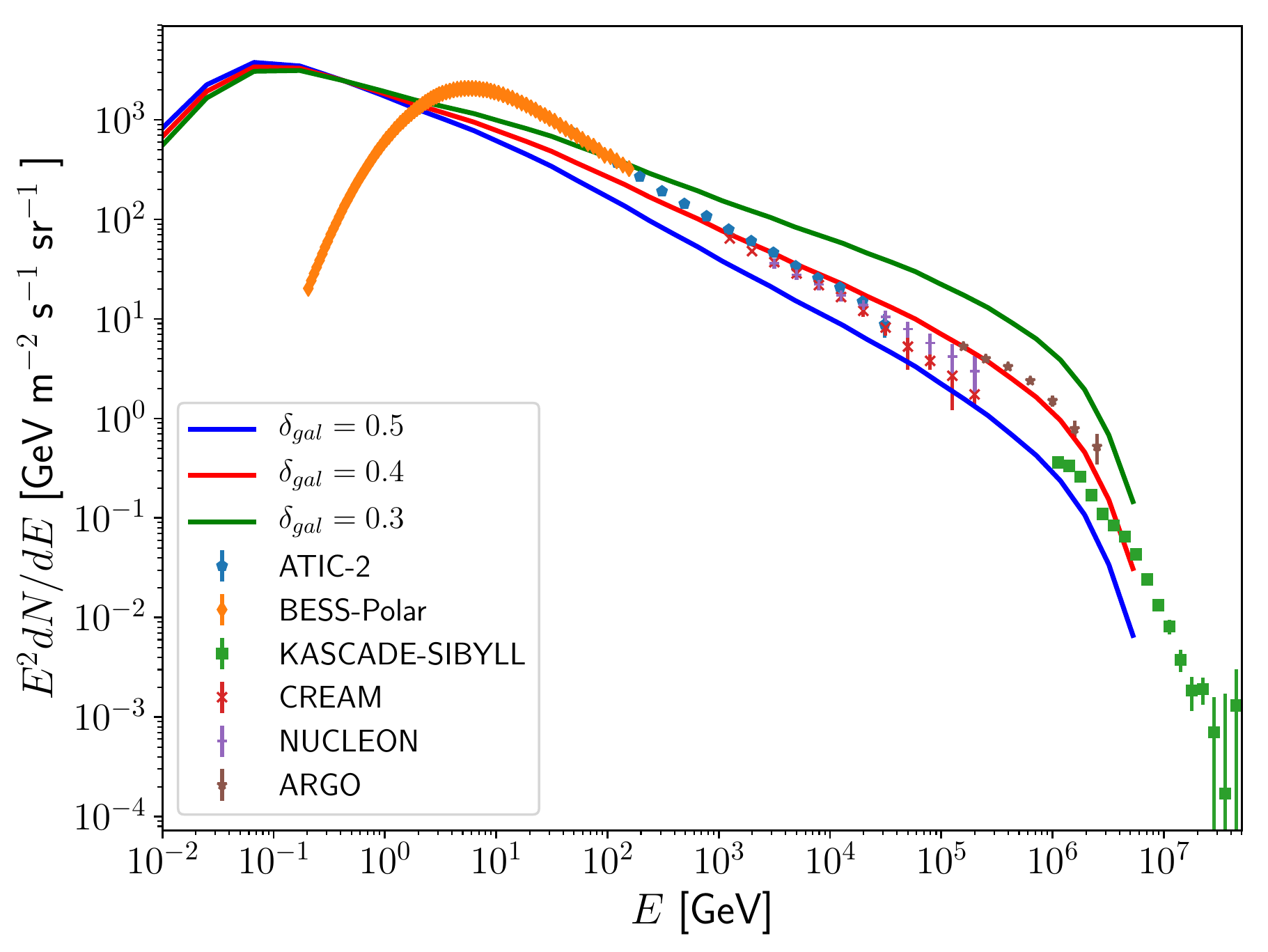}
\caption{Superbubble contribution to the galactic cosmic ray proton spectrum, assuming a galactic diffusion coefficient with index $\delta_{gal} = 0.5$ (blue), $\delta_{gal} = 0.4$ (red), $\delta_{gal} = 0.3$ (green).}
\label{fig:spectradiffuse}
\end{figure}

Let us now provide a rough estimate of the contribution of SBs enclosed by supershells to galactic CRs. The differential flux of particles reaching the Earth from SBs is estimated as:
\begin{equation}\label{CR_in_SB:diffuseemission}
\frac{\dd N}{\dd E} = \frac{f_{\rm SB}}{\langle V_{\rm SB} \rangle} \frac{T_{\rm disk}(p) c}{\tau_2} p^2 n_{\rm shell}(p) \, ,
\end{equation}
where $f_{\rm SB} \approx 20\%$ is the volume filling factor of the hot gas in the solar vicinity \citep{ferriere1998}, $\langle V_{\rm SB} \rangle \approx 0.004$~kpc$^3$ is the volume of a typical superbubble, $T_{\rm disk}(p) \approx 3 (p/{\rm GeV})^{-\delta_{\rm gal}}~{\rm Myr}$ is the residence time in the galactic disk, with $\delta_{\rm gal} \approx 0.3 - 0.5$ \citep[e.g.][]{genolini2019} and $\tau_2$ is the diffusion time in the shell given by Eq.~\ref{tau2shell}. Eq.~\ref{CR_in_SB:diffuseemission} provides an order of magnitude estimate of the flux normalisation without introducing any arbitrary parameter. Figure~\ref{fig:spectradiffuse} shows the differential flux computed from the shell energy density $n_{\rm shell}$ obtained in the previous subsection for $N_* = 100$ and a confinement parameter $\rho_b/\delta = 10$ (bottom left panel of Figure~\ref{fig:spectrashell}).

As seen in Figure~\ref{fig:spectradiffuse}, the flux normalisation is close to that of the available data. The main parameter driving the CR density in SBs is the injection efficiency at SNRs, that is the fraction of thermal particles injected in the accelerator. Increasing the injection efficiency is however not expected to change the result dramatically, for shock acceleration becomes less efficient in the nonlinear regime. The agreement between this rough prediction and the data is overall acceptable and this preliminary estimate of the contribution of SBs to the galactic CRs show that they can account for the observed CR flux with a realistic residence time and thus a realistic galactic diffusion coefficient.

Let us eventually recall that the SB contribution to the galactic CR spectrum computed beyond 1~PeV is not reliable, for we neglected the flux of particles escaping upstream of SNRs beyond the maximum energy. An accurate description of this component requires to solve the time-dependent reacceleration problem in the nonlinear regime, at least phenomenologically, which could provide an additional steep power law component at energies close to that of the knee. Such promising analysis is left for future work.

\section{Gamma-ray spectra}\label{sec:7}
The gamma-ray emission of supershells due to neutral pion decay can eventually be computed from the cosmic ray energy spectra as:
\begin{multline}
\Phi_\gamma(E_\gamma) = \frac{n_H c}{4 \upi D_S^2}
\times \\ 
\int_{T_{p,min}} \dd T_p 4 \upi p^2 n_{\rm shell} (p (T_p)) \epsilon_{nh}(T_p) \frac{\dd \sigma_{pp}}{\dd E_\gamma}(T_p,E_\gamma) \, ,
\end{multline}
where $n_H$ is the hydrogen number density of the shell, $D_S$ is the distance of the source, $T_{p,min}$ the threshold energy for neutral pion production in p-p interactions, $\dd \sigma / \dd E_\gamma$ is the differential cross-section of gamma-ray production and $\epsilon_{nh}$ is the nuclear enhancement factor accounting for nucleus-nucleus interactions. The computation follows the prescription of \citet{kafexhiu2014}.

\begin{figure*}
\centering
  \includegraphics[width=\linewidth]{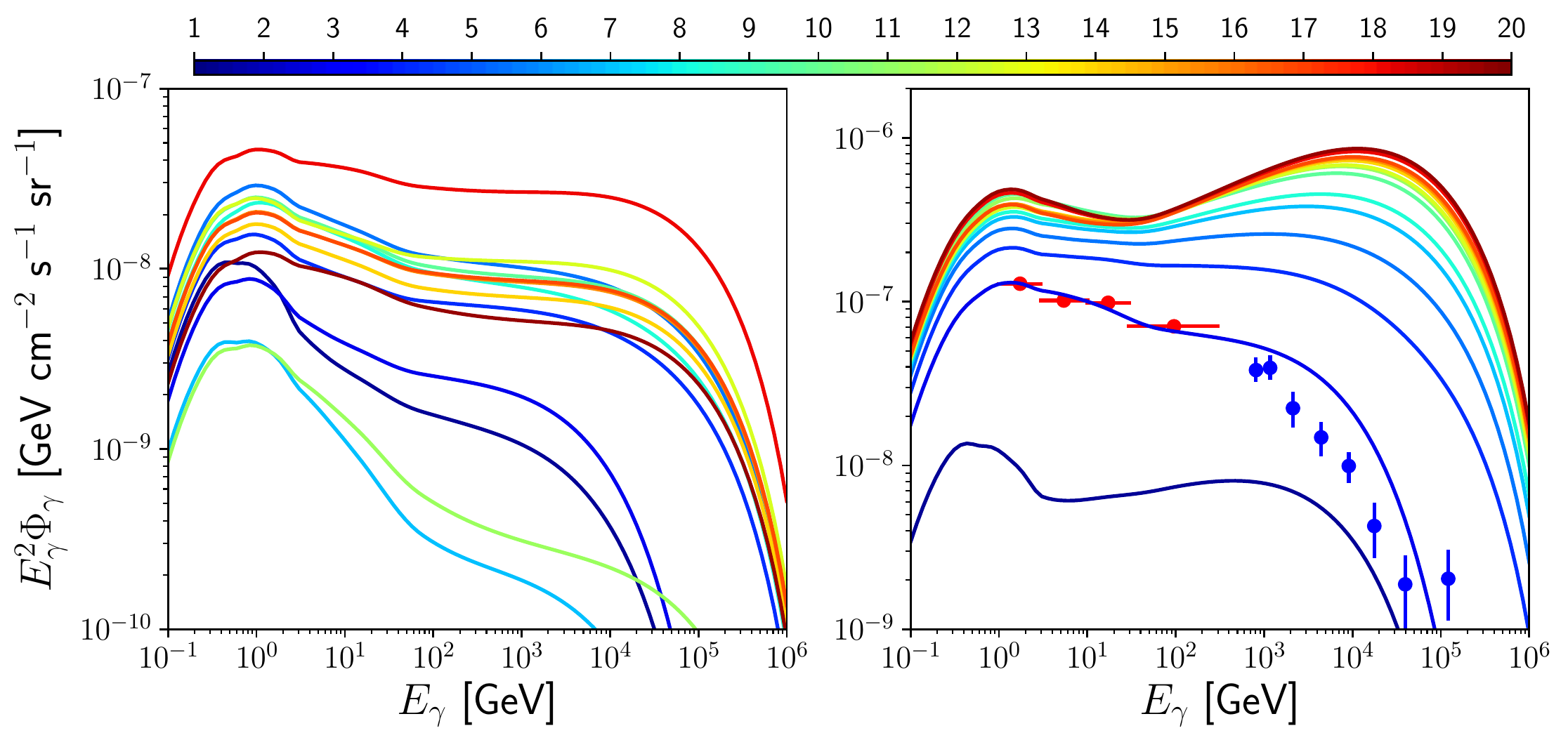}
\caption{Gamma-ray spectral energy distributions from p-p interactions in a standard superbubble shell of hydrogen number density $n_H = 40$~cm$^{-3}$ ($n_{\rm ISM} = 10$~cm$^{-3}$), with a shell confinement coefficient $\rho_b/\delta = 10$ (left panel), and 1000 (right panel). The source is a loose cluster of 100 massive stars located at 1.5~kpc from the Earth. Spectra are plotted at various times from 1 to 20~Myr corresponding to the various colours from blue to red (see the colour scale at the top). The data points show the gamma-ray flux measured in the Cygnus region: Fermi-LAT data in red \citep{abdollahi2020} and HAWC data in blue \citep{abeysekara2021}.}
\label{fig:spectragamma}
\end{figure*}

Figure~\ref{fig:spectragamma} shows the resulting gamma-ray spectral energy distributions for a cluster of 100 massive stars and two values of the parameter $\delta$. It is striking to see how the modification of a single parameter can produce many different spectral shapes. In standard SBs surrounding loose clusters, the CRs are not very efficiently reaccelerated (left panel). The fluxes escaping through the shell produce rather steep power law gamma-ray spectra, with a low normalisation and characterised by a strong intermittency. The flux is generally below the detection threshold of the Fermi-LAT telescope, which could explain why several massive star clusters are not associated with gamma-ray emissions \citep{maurin2016}, and in particular the well-known Orion-Eridani superbubble \citep{joubaud2020}. On the other hand, with an enhanced shell confinement coefficient $\rho_b/\delta$, the CRs are trapped in the shell, which results in a more efficient production of gamma-rays (right panel). The spectra at early times, driven by stellar winds, are slightly steeper than $E^{-2}$ (e.g. $\Phi_\gamma \propto E^{-2.1} - E^{-2.2}$), which is consistent with what is observed in the Cygnus-X star forming region. In particular, the spectrum at 3 Myr shows a very good agreement with the normalisation and shape of the spectral energy distribution measured in the Cygnus region, although we did not adjust any parameter to obtain the best fit. A thorough comparison with Fermi and HAWC data is left for future work. Eventually, hard power law or concave spectra are expected if the CRs are efficiently confined and reaccelerated, as it starts to be the case for $\rho_b/\delta=1000$ (right panel of Figure~\ref{fig:spectragamma}) after about $5-10$~Myr. This is expected if the perpendicular diffusion is suppressed in the shell or if the particles are efficiently reaccelerated in compact clusters. 

Several observations of extended hard gamma-ray spectra possibly associated with young massive stellar clusters or SB cavities have been reported in the last decade, even though it is not established that these gamma-ray emissions are produced by collective plasma effects, in particular because the contribution of point-like sources, e.g. pulsar wind nebulae, is difficult to isolate \citep{tibaldo2021}. Slightly concave gamma-ray spectral energy distributions were observed in the G25 region \citep{katsuta2017}, flat gamma-ray spectra were observed in the very compact and very massive Westerlund 2 young stellar cluster \citep{aharonian2007,yang2018}, and flat gamma-ray spectra extending up to TeV energies were observed in the Westerlund 1 region, in the vicinity of the most massive star cluster of our Galaxy \citep{abramowski2012,ohm2013}.
It seems that all these observations, despite differing quantitatively or even qualitatively, could be explained by SB proton spectra. Our SB modelling would allow to constrain the properties of the supershells, turbulence, and massive star clusters in order to reproduce gamma-ray data. In particular, the model being self-consistent, it predicts not only spectral shapes but realistic normalisations as well. Eventually, protons of energies above 1~PeV could produce ultra high energy photons above the sensitivity of the LHAASO observatory, which recently detected a photon of 1.4~PeV in the direction of the Cygnus OB2 massive cluster \citep{cao2021}. We however recall that the in order to accurately describe the CR spectrum produced in SBs near the maximum energy, a time-dependent nonlinear model of CR reacceleration at SNR shocks is needed.

\section{Conclusions}\label{sec:conclusions}
Using a self-consistent model accounting for all relevant ingredients rederived from first principles, we computed the acceleration of protons in superbubble environments. We extended the model of \citet{ferrand2010} in order to account for the dynamical evolution of the environment as well as the stellar winds, the losses and the effect of the shell. We also refined the model to include the backreaction of the particles onto the turbulence and onto the shocks, the latter being properly computed using an up-to-date semi-analytical model of nonlinear diffusive shock reacceleration. Our model complements the work of \citet{bykov2001} who focused on the early phase of stochastic acceleration in strong supersonic turbulence. 

We found that the stars efficiently transfer their energy into non-thermal particles by means of shock acceleration as well as stochastic acceleration in turbulence, especially around compact clusters where particles are efficiently reaccelerated in the inner region and in the case where a dense magnetised supershell prevents the particles to escape in the ISM. High CR energy densities are generally achieved in the SB, which calls for nonlinear models, and in some cases the turbulence can be completely suppressed by the non-thermal particles confined in the shell.

When the confinement is less efficient, e.g. in the limit of a thin supershell or in small clusters, the typical spectra are rather intermittent, displaying a typical ``Fermi II bump'' from the injection energy to the GeV band, and then transitions toward a steep power law produced by the nearly stationary wind contribution as well as the intermittent supernovae modulated by the escape. Providing the level of turbulence is about a few percent, which requires the stars to transfer a few tens of percent of their mechanical power into hydromagnetic waves, the low energy particles are efficiently reaccelerated, which gives
rise to steep escape fluxes typically scaling as $E^{-2.2}$, which is close to what is needed in order to account for the diffuse CR spectrum observed near Earth. When the confinement of the particles is enhanced, the spectra harden with typically flat gamma-ray signatures, or even small concavities with somewhat hard components (about $E^{-1.8}$ at high energies) due to the successive reaccelerations of the confined particles.

The variability of the spectra from cluster to cluster, and also during the lifetime of a given cluster, provides a simple answer to the puzzling discrepancies between the recent gamma-ray observations of SBs and star clusters. Indeed, some SBs are not detected in gamma-rays (e.g. the Orion-Eridani SB or the Rosette nebula), others display rather steep power law spectra (e.g. the Cygnus region), or flat energy distributions (e.g. the Westerlund 1 and 2 regions), and some data even suggest slight concavities (e.g. in the G25 region). The specific gamma-ray signature of a SB can therefore be used to constrain the properties of a given massive star cluster (its number of massive stars, if it is compact or loose...), the surrounding environment (the magnetic field, the turbulence level...), as well as the properties of the supershell (the thickness, the density...). Although the model depends on several parameters, some of them such as the cluster and shell properties can hopefully be constrained by multiwavelength observations. The main difficulty is probably to infer the properties of the turbulence (intensity and spectrum), which at the moment are unknown. Using our model, gamma-ray spectra could be used indirectly to constrain the diffusion coefficients. Indeed, the momentum diffusion coefficient drives the CR spectrum slope at low energies by means of the stochastic reacceleration, while the spatial diffusion coefficient drives the CR spectrum slope at high energies as well as the intensity of the gamma-ray flux, because it determines the efficiency of the confinement. Such analysis could be applied for example to the Cygnus region.

Eventually, the overall SB contribution to the galactic CR population can be estimated using Monte-Carlo samplings. Thorough statistical computations confronted to the observed CR spectrum could be used to probe the most likely SB parameters, in particular the magnetic fields, turbulence levels and shell properties, as well as their variance. The modulation of the galactic spectrum due to the local bubble could also be included in such computation using our two-zone modelling. Indeed, this model allows to compute the screening of the interstellar CR flux when it penetrates a dense supershell, an idea which was already raised in the 80's (\citealt{streimatter1985}, see also \citealt{bouyahiaoui2019,silsbee2019,phan2020}).


\section*{Acknowledgements}
TV acknowledges A. Marcowith, E. Parizot, V.H.M. Phan, L.M. Bourguinat for helpful discussions and suggestions. SG and VT acknowledge support from Agence Nationale de la Recherche (grant ANR-17-CE31-0014).


\section*{Data Availability}
No new data were generated or analysed in support of this research.



\bibliographystyle{mnras}
\bibliography{biblio} 


\appendix
\section{Notations}\label{app1}
Table~\ref{tab:notations} provides an exhaustive list of the notations used throughout the present paper.

\begin{table*}
\footnotesize
\centering
\begin{tabular}{lcr}
\hline
\textbf{CLUSTER AND SUPERBUBBLE} & & \\
*Initial number of massive ($M>8 M_\odot$) stars & $N_*$ & $100 - 1000$ \\
Lifetime of the superbubble & $\mathcal{T}_{\rm SB}$ & 35 Myr \\
Time-dependent mechanical power & $\mathcal{P}_{\rm tot}$ & $10^{38}-10^{39}$ erg/s \\
Average mechanical power (winds and supernovae) & $L_*$ & $10^{38}-10^{39}$ erg/s \\
Fraction of the mechanical power working on the shell & $\xi_b$ & 22\% \\
Radius of the superbubble & $R_b$ & $50 - 150$~pc \\
Superbubble radius normalised per star at 1 Myr & $R_0$ & 10 pc \\
Volume of the superbubble & $V_{\rm SB}$ & \\
*Density of the interstellar medium around the superbubble & $n_{\rm ISM}$ & $10-100$ cm$^{-3}$ \\
Interior number density & $n$ & $0.01 - 0.1$ cm$^{-3}$ \\
Interior density & $\rho$ & $0.2 - 2 \times 10^{25}$ g cm$^{-3}$ \\
Interior temperature & $T$ & $5 \times 10^6$ K \\
\hline
\textbf{MASSIVE STAR WINDS} & & \\
Wind mechanical power & $\mathcal{P}_w$ & $10^{37} - 10^{38}$ erg/s \\
Radius of the wind termination shock & $R_s$ & $1 - 10$ pc \\
Velocity of the wind outflows & $V_w$ & 2000 km/s \\
*Injection momentum at winds & $p_0$ & 5 MeV/c \\
*Maximum momentum at winds & $p_w$ & 10$^5$ GeV/c \\
*Wind acceleration efficiency & $\eta_w$ & 10\% \\
\hline
\textbf{SUPERNOVA REMNANTS (SNR)} & & \\
Supernova energy & $E_{\rm SN}$ & $10^{51}$ erg \\
Mass of supernova ejecta & $M_e$ & $10~M_\odot$ \\
Initial velocity of supernova shocks & $V_{\rm SN}$ & 3000 km/s \\
Velocity of SNR starting the Sedov-Taylor phase & $u_0$ & 1500 km/s \\
``Equivalent'' volume spanned by ``stationary'' SNR & $\mathcal{V}_{\rm SNR}$ & 1\% $V_{\rm SB}$ \\
Fraction of cosmic rays reaccelerated at SNR & $\chi$ & 1\% \\
*Injection parameter of SNR shocks & $\xi$ & 3.5 \\
Maximum momentum in SNR shocks & $p_{\rm max}$ & 1 PeV/c \\
\hline
\textbf{TURBULENCE} & & \\
Source of the turbulence at the largest turbulent scale & $S$ & \\
Energy spectrum of the turbulence & $W$ & \\
*Efficiency of turbulence generation & $\eta_T$ & $1 - 30$\% \\
*Relaxation time of the turbulence generated by SNR & $\tau_T$ & 1 Myr \\
*Largest turbulent scale & $\lambda$ & 10 pc \\
Smallest turbulent wavenumber & $k_0$ & \\
Non-thermal turbulence damping rate & $\Gamma$ & \\
*Large scale magnetic field & $B_0$ & 10 \textmu G \\
Random component of the magnetic field & $\delta B$ & $1 - 10$ \textmu G \\
Total magnetic field & $B$ & $10 - 20$ \textmu G \\
Turbulence level & $\eta$ & $1 - 10$\% \\
Alfvén velocity & $v_A$ & $10 - 100$ km/s \\
Root-mean-square velocity & $\delta u$ & $10 - 100$ km/s \\
\hline
\textbf{PARTICLE TRANSPORT} & & \\
Particle distribution function & $f(x,p,t)$ & \\
Particle spectrum & $n(p,t)$ & \\
Particle velocity & $v$ & $0.3 - 3 \times 10^8$ m/s \\
Particle kinetic energy & $\epsilon(p)$ & 10 MeV $-$ 10 PeV \\
Spatial diffusion coefficient & $D_x$ & $10^{27-28}$ cm$^2$/s at 1 GeV/c \\ 
Momentum diffusion coefficient & $D_p$ & 1 (GeV/c)$^2$/Myr at 1 GeV/c \\
Escape time from the superbubble & $\tau$ & 1 Myr at 1 GeV/c \\
Flux of particles coming from the interstellar medium & $\phi_{\rm ISM}$ & \\
Flux of particles escaping in the interstellar medium & $ \Phi(E) $ & \\
\hline
\end{tabular}
\caption{Notations used in the present work. The third colum provides typical values. The input parameters of the model are highlighted with a star.}
\label{tab:notations}
\end{table*}


\bsp	
\label{lastpage}
\end{document}